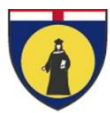

# Multifunctional nanostructures for intracellular delivery and sensing in electrogenic cells

*Giulia Bruno*

*Thesis submitted for the degree of Doctor of Philosophy (XXXII° cycle)*

**Thesis Jury**

*Patrizio Candeloro*

*Francesco Gentile*

**Supervisor**

*Francesco De Angelis*

**Head of PhD**

*Giorgio Cannata*

Dibris

Department of Informatics, Bioengineering, Robotics and Systems Engineering

# Contents

# Summary

Biological studies on *in vitro* cell cultures are of fundamental importance for investigating cell response to external stimuli, such drugs for specific treatments, or for studying communication between cells. In the electrophysiology field, multielectrode array devices (MEA) are the gold standard for the study of large ensambles of electrogenic cells. Thus, their improvement is a central topic nowadays in neuroscience and cardiology [1]. In the last decades, thanks to the adoption of nanotechnologies, the study of physiological and pathological conditions of electro-active cells in culture have becomes increasingly accurate[2], allowing for monitoring action potentials from many cells simultaneously.

In fact, nanoscale biomaterials were able to overcome the limitations of previous technologies, paving the way to the development of platforms for interfacing the electrogenic cells at unprecedented spatiotemporal scales. These devices, together with microfluidics, are starting to be used for drug screening and pharmaceutical drug development since they represent a powerful tool for monitoring cell response when cultures are stimulated by target compounds. Many pharmaceutical agents, however, including various large molecules (enzymes, proteins, antibodies) and even drug-loaded pharmaceutical nanocarriers, need to be delivered intracellularly to exercise their therapeutic action inside the cytoplasm[3]. Nanoscale electrodes offer individual cell access and non-destructive poration of the cellular membrane enabling high capability in the delivery of biomolecules. Among all the techniques, electroporation have proven encouraging potential as alternative to the carrier mediated methods for molecular delivery into cultured cells[4].

In this regard, different groups [5][6][7] exploited the integration of nanostructures with delivering capabilities with single-cell specificity and high throughput in biosensing platforms. These efforts provided powerful tools for advancing applications in therapeutics, diagnostics, and drug discovery, in order to reach an efficient and localized delivery on a chip.

Despite these new tactics, there is still a critical need for the development of a functional approach that combines recording capabilities of nanostructured biosensors with intracellular delivery. The device should provide for tight contact between cells and electrode so as to enable highly localized delivery and optimal recording of action potentials in order to attain a high degree of prediction for the disease modeling and drug discovery. This "on-chip" approach will help to gain deeper insight in several bio-related studies and analyses, providing a comprehensive knowledge of the entire cellular dynamics when selectively stimulated by the desired bio-molecules.

In the first part of this dissertation, a solution will be proposed in order to fill this gap and respond to this need in the biology field.

In the first chapter, I will describe briefly the principles of action potentials and how neurons and cardiomyocyte are composed, together with the development of electrophysiology and the advent of multielectrode arrays.

In the second chapter, more details about fabrication and cell-electrode system modelling will be explained. In the same chapter, I will explore the development of multielectrode arrays up to the present days, along with the advent of nanotechnologies and the related techniques for improving the previous platforms. The different cell poration techniques will be described in order to reach the best recording capabilities without damaging cells. Electroporation, optoporation and spontaneous poration will be presented and the chosen technique for our application (electroporation) will be reviewed more in detail.

In the third chapter, different methodologies for intracellular delivery will be explained, focusing also on the electroporation technique. A small paragraph about the integration of these techniques on chip will be inserted to illustrate the state of the art of these devices.

The fourth chapter will explicate in details the **Microfluidic multielectrode array** idea, the approach used in order to fabricate this novel platform from scratch, the experiments carried out to verify its capabilities and the associated results.

In the last paragraph, I will discuss how the proposed platform could became suitable for the day to day uses in research activity by employing nanoporous materials.

In fact, big efforts are carried out in order to find appropriate metamaterials as substitutes of the 3D counterparts so as to decrease the cost of device manufacturing that makes them unfitting with research activity.

As a novel electrode material, nanoporous metals possess unique properties, such as a low fabrication cost, high plasmonic enhancement and large surface-volume ratio[8].

Nanoporous gold behaves like a metamaterial whose effective dielectric response can be tuned accordingly to the wanted use. These properties make the material suitable for multiple biosensing application, from a high-performance and reliable SERS (surface enhanced raman scattering) substrate [9] to an electrode in CMOS MEAs capable of intracellular recordings[10][11][12].

All these properties were explored in the last years, but it could be interesting to further study if the characteristics of this material could make it a good photoelectrical modulating material for eliciting electrogenic cells firing activity. In this way, this technology could be in principle easily implemented on commercial CMOS devices, consenting stimulation and recording at single cell level with high-resolution sensors, opening the way to new methodologies for studying electrogenic cells and tissues.

Electrical stimulation of excitable cells is the basis for many implantable devices in cardiac treatment and in neurological studies for treating debilitating neurological syndromes. In order to make the technique less invasive, optical stimulation was widely investigated [13]. The non-genetic photostimulation is starting to make its way in the field since it allows to avoid changing the biological framework by using transient thermal or electrochemical outputs from synthetic materials attached to the target cells[14]. If stimulated with impinging light these materials could inject free charges into the solution resulting in an ionic current at the interface able to eliciting of neurons[15] or cardiomyocyte action potentials.

Plasmonic porous materials have all the suitable properties to be considered as an appealing tools for charge injection and consequently for stimulation of electrically active cells [16].

Thus, the second part of this dissertation will exploit the capabilities of these plasmonic metamaterials, placing particular emphasis on the possibility of photoelectrochemical modulation.

In particular, in the fifth and last chapter I will describe all the properties and application of the porous material and the mechanism of photoemission.

In the experimental paragraphs, the free charge photoemission properties of porous gold will be explored together with **plasmonic non-genetic photostimulation** of the cardiac cells on commercial CMOS MEAs.

# 1 Cell signalling and electrophysiology development

## 1.1 Basics of cell communication and its main components

### *1.1.1 Action potential*

The action potential is the method by which electrogenic cells such as cardiomyocytes or neurons manage to communicate with one another. This potential consists in a rapid depolarization and repolarization of a limited area of the membrane caused by the opening and closing of the ion channels[17]. Action potentials can travel by meters without any loss of signal strength, because it is regenerated continuously (for example along the axons in neurons). The whole process is regulated by the presence of voltage regulated access ion channels located in the membrane. The electrical polarization results from a complex interplay between protein structures embedded in the membrane called ion pumps and ion channels. It is important to define two relevant levels of membrane potential: the resting potential (typically –70 mV ), which is the value maintained by the membrane potential as long as the cell is not perturbed, and a higher value called the threshold potential (about −55 mV for neural activity).

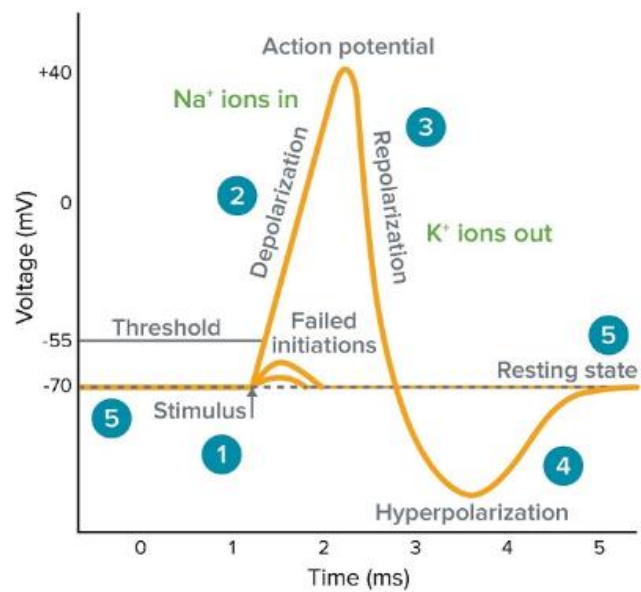

**Figure 1** *Schematic drawing of action potential. Image from Molecular Device*

These electrical properties are determined by the structure of the membrane that surrounds the cell. In fact, the voltages inside and outside the neuron are different because the membrane is more permeable to K+ than it is to Na +. The strength of the concentration gradient pushes a certain amount of $K^+$ out of the cell, thus leaving slightly more negative inside the neuron than its exterior.

In the first part of the event, when the threshold is reached, the sodium voltage-gated ion channels are opened and this type of ions flows quickly into the cell. This influx further depolarizes the cell, opening other channels for sodium, continuing in this way the self-reinforcing cycle, in a process called the Hodgking-Huxley cycle. If enough ions get inside, the event takes place otherwise it will be a failed initiation. In the positive case, the membrane will depolarize up to sodium Nerst equilibrium potential (+63 mV ). In this phase, $Na^+$ channels inactivate while the potassium voltage-gated ion channels open up allowing these ions to flow outwards following their concentration gradient. This outflow leads to the return of the potential with the one of the membrane. The opening of these latter channels lasts longer than that for sodium, causing a repolarization phase. This process pushes the potential towards the potassium equilibrium potential which is more negative than the resting potential. As a result the membrane is temporarily hyperpolarized (-80 mV). This hyperpolarization causes the closure of the potassium channels, and therefore a gradual return to resting potential. During this transient state of hyperpolarization, the sodium channels cannot reopen and therefore no further action potential can arise. This phenomenon is called the absolute refractory period. This is followed by a relative refractory period along which a neuron can produce an action potential only following strong depolarization currents. The period lasts only a couple of milliseconds.

## 1.1.2 *Neuron*

Nervous system is composed by two main kinds of cells: neuron and glial cells. Neurons are the fundamental building blocks for elaborating signals and send the information in all the nervous system. Neurons are distinguished by shape, function, location and interconnection within the nervous system. Glial cells are considered as non-nervous cells that perform different functions, such as furnishing structural support, isolating electrically neurons and modulating neural activity[17][18].

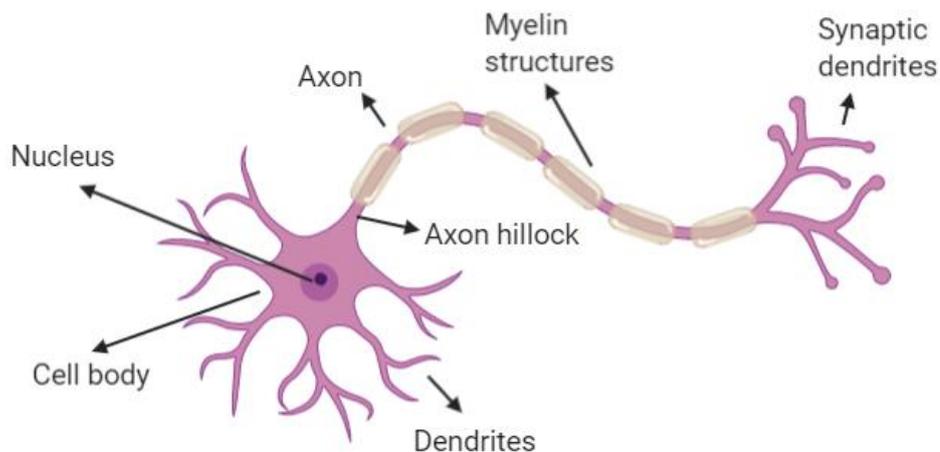

**Figure 2** *Schematic of a neuron with main components*

Such as in the main number of eukaryotic cells, even in the neurons it is possible to find the standard cellular components. The cellular membrane encloses the cells body, which it is called soma. This structure is the metabolic apparatus that sustain the vital functions of the cells. Nucleus, endoplasmic reticulum, cytoskeleton, mitochondria, Golgi apparatus and other organelles are enclosed in the body. The other predominant component of these cells are dendrites and axons. The first are branched extensions of the neuron that receive input from other neurons. They can take many different and complex forms, depending on the type and location of the neuron. They may appear as large arborizations, similar to the set of branches of a tree (such as Purkinje cells in the cerebellum) or they may take simpler forms such as dendrites of spinal motor neurons. On the dendrites are present also specialized cytoplasmic processes called thorns. These latter are small protuberances fixed on the surface of the dendrites, where the dendrites receive input from other neurons. The axon, instead, consists of a single process that extends from the cellular body. This structure represents the output of the neuron, along which the electrical signals descend to the axonal endings, where the neuron transmits the signal to other neurons or to other types of cells. Transmission occurs in synapses, a specialized structure where neurons are in close contact among them, so that chemical or electrical signals can be passed from one cell to another. Some axons branch out to form collateral axons that can transmit signals to more than one cell. Many axons are wrapped in a insulating substance called myelin. Along the length of the axons, there are uniformly spaced breaks in the myelin structure, that it is called Ranvier nodes.

The processing and transmission of the neural signal can take place within the neuron itself or between adjacent neurons. Within the neuron, the information transfer involves modifications of the electrical conduction of the neuron while the signal flows along its volume. Between neurons, instead, the transfer occurs in synapses, generally mediated by chemical transmission molecules (neurotransmitters) or in some cases by real electrical signals.

### *1.1.3 Cardiomyocyte*

Cardiomyocytes carry out the contractile function of the heart that enables cardiac muscles to pump blood throughout the body. This is achieved through an excitation-contraction coupling that converts action potentials into muscle contraction. Examinations reveal that this cell are equipped with striated myofibrils, interspersed with rows of mitochondria; essential for contraction. The myofibrils contain two types of myofilament, thin filaments of actin and thick filaments of myosin.

During an action potential, membrane depolarization results in an influx of calcium ions into the cell. As the calcium binds to receptors inside the cell, this results in the release of even more calcium into the cell (through calcium channels). At the same time, adenosine triphosphate (ATP) is supplied by the mitochondria situated alongside the myofibrils. With the energy obtained by ATP, the head portion of myosin binds to actin resulting in muscle contraction. In fact, the effective contraction is given by the molecular structure of myosin, which has a series of heads that project outwards to interact with the thin filaments and act as ATP binding sites and the regulatory proteins, troponin and α-tropomyosin, which are associated with the actin and confer calcium sensitivity [19]. In turn, this outcomes in the shortening of actin-myosin fibrils in the cell.

## 1.2 History of electrophysiology and MEA development

Microelectrodes have been extensively used to investigate the electrophysiological activity of the nervous system for more than a century. The foundation of the field, could be considered dated back to the famous Galvani's work, published in 1791. In his book, *De Viribus Electricitatis in Motu Musculari,* he described the discovery of the electrical excitation of the nerve–muscle preparation using an electrical machine; the Leyden jar (nowadays could be defined as a capacitor). Several years later Galvani made another fundamentally important experiment demonstrating for the first time the concept of propagating bioelectricity. This time he connected the two frog legs along the sciatic nerve and when the nerve of the first preparation was in contact with the second one, a contraction happened in both preparation. The controversy and critical experiments conducted by Alessandro Volta, opened up the field to a wide number of scientist that investigated the phenomena of 'animal electricity'. Caton, Bernstein, Einthoven, Hodgkin and Huxley, all together, laid the foundations of the modern study of electrophysiology [20].

The first recording of electrical activity from neuromuscular preparation was conducted by Nobili using an electromagnetic galvanometer, developed in 1825. He measured a current associated with muscle contraction which he called *corrente di rana* (frog's current). Two decades passed before the right interpretation was conceived when in 1840s Johannes Peter

Müller developed the concept of electrical signal propagation though the nerve. Nevertheless, the first machine that was able to record resting and action potential was the 'differential rheotome' invented by Bernstein[21]. This machine was able to measure resting and action potential for the first time allowing for the estimation of the resting potential at ~ -60 mV and to develop the membrane theory of excitation.

The direct measurement of ion-carried transmembrane currents was prompted by the discovery of the giant squid axon by John Z. Young in 1936 [22]. This led to the validation of the ionic theory of membrane potential and membrane excitability. Using the squid axon model, Cole and Curtis measured with extracellular electrodes the changing in the impedance, revealing the rapid decrease in membrane resistance during the propagation of the action potential, which was indicative of transmembrane current generation. Successively, Cole and Curtis [23] and independently Hodgkin and Huxley [24] fabricated mini-electrodes which could be inserted into the axon and preformed first direct measure of resting and action potentials. Several years later Cole and Marmot developed the voltage clamp technique that allowed for direct recordings of membrane currents[25].

From that moment on, the way for manufacturing the most appropriate tools for exploring to the fullest the properties of bioelectrical signalling of bio-entities was paved by the efforts of many researchers and laboratories. The beginning of single cell electrophysiology is directly associated with the introduction of glass microelectrodes with micron tips suitable for low-traumatizing penetration of individual cells, developed by Gilbert N. Ling and Ralf W. Gerard in 1949[26]. The very first direct recordings of currents associated with openings of ion channels in artificial lipid bilayers were made by Paul Müller and Donald O. Rudin in 1962[27]. E. Neher and B. Sakmann developed the patch clamp technique in between late 70s and 80s that earned them the Nobel Prize in 1991[28]. This technique, for the first time, allow to record the currents from single ion channel, increasing the knowledge about the involvement of ion channels in fundamental cell processes such as action potentials.

In parallel to single cells recording, the multiple extracellular electrodes method was developed to respond to the requirements of investigating the collective behaviour of cultured cells. In 1972, Thomas et al. published the first paper describing a planar multielectrode array for use in recording from cultured cells[29]. The multielectrode array (MEA) consisting in two rows of 15 electrodes each, spaced by 100 µm, was intended as a new platform for studying cultured cardiac myocytes. Few years later, the first successful recordings from single dissociated neurons were reported by Pine from a multielectrode array with two parallel lines of 16 gold electrodes and insulated with silicon dioxide [30]. Following this trend, Gross reported successful experiments with cultured ganglion cells [31]. In 1986, Wheeler and Novak reported the measurement of extracellular field potentials from brain slices and flexible MEAs were developed for elongated slice experiments[32].

In the next paragraphs will be explained in details standard MEA fabrication techniques and the equivalent electrical circuit that models the cell-electrode system in order better understand the phenomena at the interface.

# 2 Multielectrode arrays and nanotechnologies

## 2.1 Fabrication techniques and theoretical model

### 2.1.1  Standard MEA fabrication

Since the MEA was first introduced, it has been gradually improved by the newly developed microfabrication techniques[33][34][35]. To fabricate multi electrode array, the materials for the substrate, the microelectrodes and insulation layers should be carefully selected. The choice of materials is strictly related to the intended use of the devices. Other important aspects have to be considered, such as, the biocompatibility, the transparency and durability in cell culture conditions.

The benchmark materials for metal sensing pads are gold, platinum, titanium nitride and indium-tin oxide (ITO). The metal electrodes and conductive feedlines are made through micropatterning techniques and the number of the electrodes is decided by the pattern design. In details, the desired pattern is made through photolithography that leads to a change in the solubility of the desired area of the material and the consequent immersion in a developer solution will dissolve away the soluble areas of the photoresist. Secondly, a thin metal layer is deposited onto the patterned substrate (silicon or glass) using sputtering, thermal evaporation or e-beam evaporation. In case of gold or platinum, an additional metal layer such as titanium is required to promote the adhesion between the substrate and metal and to prevent peeling while the device host the cell cultures.  As the last step, feedlines need to be passivated by an insulator layer. Since the device is working in aqueous conditions containing abundant ions, a good insulator should minimize signal crosstalk and signal attenuation. Silicon dioxide, silicon nitride, polydimethylsiloxane, SU-8  or other polymeric layers are the most used for the purpose.

To read electrical signals at the microelectrode, MEAs need to be connected with an external read-out circuitry. The contact pads could be directly connected with the amplifiers or alternatively, each chip could be package with a printed-circuit board adaptor that connects the MEA chip with the circuitry. The connection between contact pads and the PCB board is realised by wire-bonding or conductive glues. At last, in order to grow cells on the platform, a glass or Teflon ring will be attached to  hold cell culture medium.

## 2.1.2 Equivalent circuit model

### 2.1.2.1 Cell- electrode model

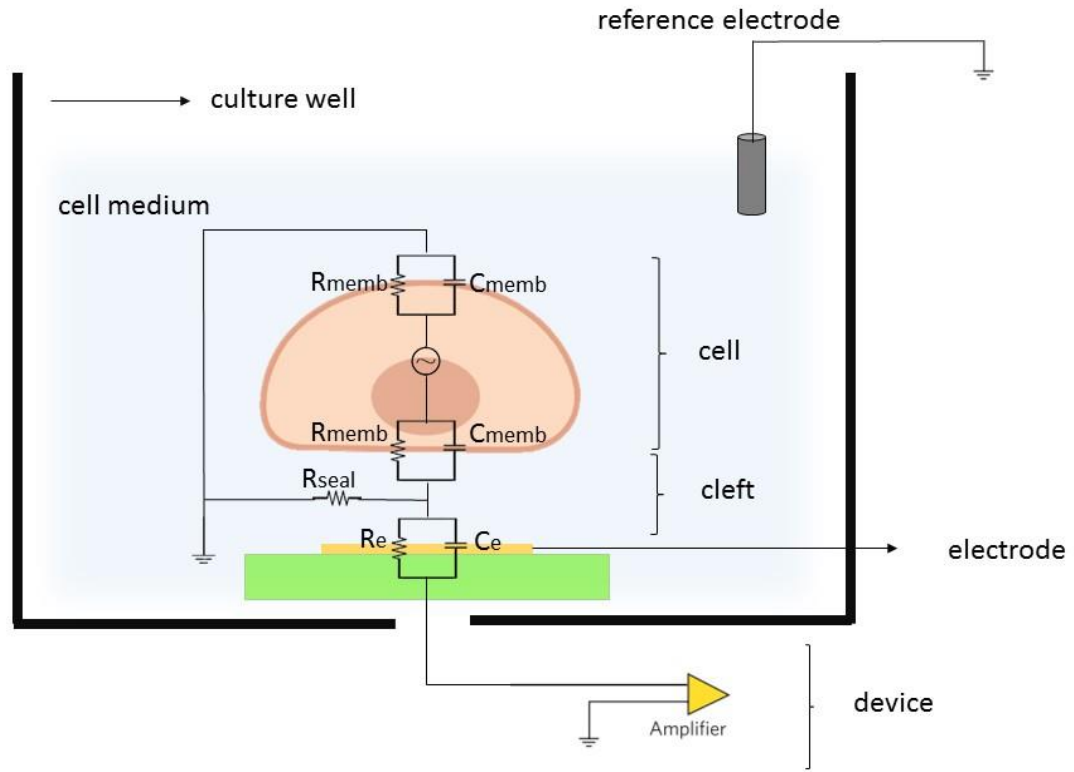

**Figure 3** *The device is represented by RC$_{electrode}$ (green substrate, gold electrode), the cleft is the sealing resistance and the cell is the current generator while the membrane is represented by RC$_{membrane}$. All together they form an equivalent electrical model representative for cell electrode system, including the phenomena at the interface between cell and electrode.*

When electrogenic cells, such as neurons or cardiomycytes are cultured on a MEA, they adhere to the surface of the platform and make direct contact with microelectrodes. MEA devices record extracellular action potentials from the membrane of the cell upon the electrode. The recording principles are well explained by this simplified electrical circuit model that includes cells and electrode (fig 3). In this model, the cell membrane, consisting of a double lipid layer spanned with protein channels, is conceived as a RC circuit with a current generator. In fact, the lipids of the cell membrane act as a capacitor, while the channels allow a slow passage of ions functioning as a resistor. This part of the circuit deform the recorded signals, which will be different in shape from the real intracellular action potentials. In fact, the RC circuit acts as a differentiator circuit, and the output signal results in the typical biphasic shape of the field potential. The cleft formed between the cell and the MEA electrode is filled with the ionic solution that is represented with a resistor. So, when action potentials are generated from cells, it creates extracellular voltage (field potential) across the seal resistance. Thus, microelectrodes can detect the field potentials, which can span from tens to hundreds of microvolts to few mV.

By artificially increasing the seal resistance, it is possible to avoid leakage into the cell medium and thus to record higher quality potentials. For these reasons, demanding efforts were done in order to increase the value of $R_{seal}$. Studies of cell/electrode interfaces showed typical cleft thicknesses between 40–100 nm. Different studies were made in this direction in order to understand the phenomena at the interface and to comprehend how to improve this coupling[36].

Moreover, in the system depicted in fig. 3 it is possible to notice two electrodes: reference electrode and working electrode. The latter is defined as the electrode under examination, while the first is necessary to complete the circuit for charge conduction and to define a reference in case of electrical potential measurements. The electrode is coupled with an amplifier that allows for filtering the analogic signal. The shape and amplitude of the signals recorded by the electrodes is also attributable to the transfer properties of the electrical impedance generated by the ionic bilayer formed at the interface between the electrodes and the culture medium and of the AC amplifier used. Another factor that widely affects the electrical coupling coefficient between cells and MEAs is, indeed, the input impedance of the electrode.

### 2.1.2.2 Electrical double layer

In order to well explain the modelled circuit and the impedance at the electrode it is important to introduce the physics at the interface among the electrode and the cell medium.
When a conductor, such as the electrode (depicted in fig.4) is placed in contact with an electrolyte (for instance the physiological medium), a structure appears among the two. In the very narrow interface between the electrolyte and electrode a change in electrical potential occurs even in equilibrium condition, when no current is applied. This interface is called, the electrical double layer and it refers to two parallel layers of charge surrounding the object.

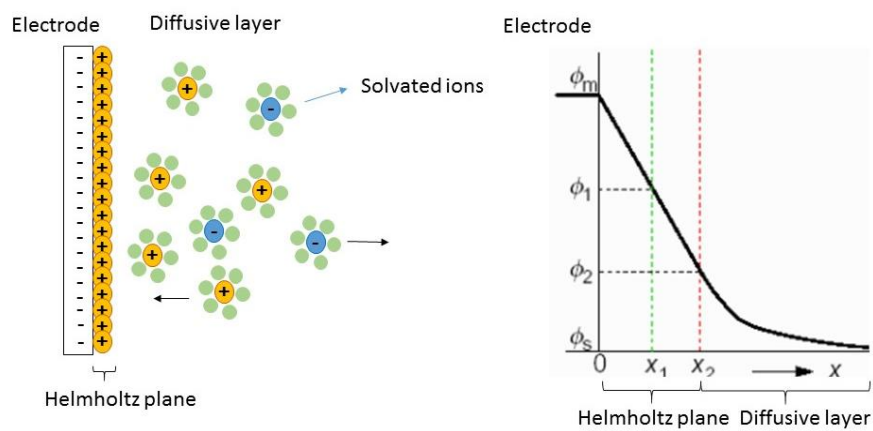

**Figure 4** *Schematic of the charge distribution in the double layer and (right) potential trend with respect to the distance from the electrode*

The inner layer, the closest to the electrode, contains solvent molecules and sometimes other species (ions or molecules) that are specifically adsorbed. This inner layer is also called the compact Helmholtz (or Stern) plane and it is due to chemical interaction. The second layer is composed of solvated ions attracted to the surface charge via long-range electrostatic forces. This second layer, also called diffusive layer, is made of free ions that move in the fluid under the influence of electric attraction and thermal agitation of the solution. The thickness of the diffuse layer depends on the total ionic concentration in the solution; for concentrations larger than 10 mM, the thickness is less than 100 Å.

Hermann von Helmholtz was the first to demonstrate that charged electrodes immersed in electrolytic solutions accumulate charge in the form of ions on their surface [37]. From that moment on, different mathematical models were proposed to explain the phenomena. The accepted model involves a linear potential drop in the compact layer and an exponential decrease with respect to the distance in the diffusive layer. Hence, given the above description, the Helmholtz model is equivalent in nature to an electrical capacitor with two separated plates of charge, for which a linear potential drop is observed at increasing distance from the plates. The electrode could be modelled as a capacitor (due to the double layer) and a resistor (due to the electrochemical reactions at the electrode) resulting in the RC circuit shown in fig. 4.

## 2.2 Multi electrode array from past to present

Since the early pioneering works, this method has become the prime methodology for studying neuronal cells connectivity, physiology and pathology in *in vitro* or *in vivo* conditions on the network level. The MEA technology has been applied to neural network studies, electrogenic cells analysis and drug screening studies[1][38][39]. The reason of this success resides in its unique features. On one hand, the multiple electrodes on the surface give to the MEA technology the ability to provide a spatiotemporal measurement platform, enabling the simultaneous recording and stimulation at different sites. Second, MEAs provide a non-invasive cell-electrode interface that allows long-term recording and stimulation for days and months without inflicting mechanical damage to the cells. The cultured cells can maintain electrical activities for more than a month being in a controlled culture environment[40].

Extracellular field potentials (FP) recorded by the MEA devices reflect the spike activity of electrogenic cells or the superposition of fast action potential (AP) and synaptic potentials in both time and space. Even though a lot of information can be gained by using multiple electrodes, the information extrapolated from spike-pattern features is limited. For example, the collective behaviour of the network could be clearly investigated by the device but this extensive spike sorting cannot provide single cell resolved information. The device is blind to resolve where the excitatory inputs by endogenous mechanisms or hyper-polarization of the membrane potential happen in a single neuron. Undetectable by extracellular electrodes are also subthreshold potentials, that mediate significant part of neuronal signalling[1].

The patch-clamp technique[41], on the contrary, is valued for its highly sensitive intracellular recording that can record propagation of APs in neurons and also subthreshold events such as postsynaptic potentials. The method is regarded as the gold standard for ion channel research, since it offers direct insights into ion channel properties through the characterization of ion channel activity[42]. Despite these qualities, the major drawbacks regard the low screen throughput and high invasiveness. Moreover, the use of sharp or patch microelectrodes is limited to individual neurons and the inset of the electrode tips into target cells requires the use of bulky micro-manipulators. Furthermore, the duration of intracellular recording sessions is limited by mechanical and biophysical instabilities. All these reasons make the technique impracticable for evaluating network behaviour in a large number of cells.

An ideal multifunctional readout system should provide information that cover the entire repertoire of electrophysiological parameters from the individually recorded neurons to the network interactions between the single units. As the technology evolved, there have been various engineering solutions to answer to the needs in diverse application areas of MEAs. The advent on nanoscale technologies changed the way of perceiving multielectrode arrays structures and application. In fact, using nano- and micro-technologies, a number of laboratories began to merge the advantages of substrate-integrated extracellular MEA technologies with the critical advantages of intracellular electrodes.

## 2.3 Advent of nanotechnology

Nanoscale biomaterials can overcome the limitations of MEA technologies, representing a powerful tool for interfacing the nervous system at unprecedented scales [2]. First and foremost, the use of nanostructured electrodes means that the electrodes present very low impedance, improving the resolution of current recording and modulatory systems. In fact, nanostructures such as porous platinum, nanopillars or nanoflakes allow increasing the surface that interface the cell compensating for the dimensions of the electrode. In fact, the cellular membrane bends around the nanostructures following the topography of the nanostructured electrodes. This behaviour is widely studied in the field through multiple techniques, and it has been demonstrated to improve the coupling between cell and electrode leading to a tight engulfment of the electrode. The resistance between cell and electrode ($R_{seal}$) is in this way brought to very high value decreasing leakage in current across the cellular medium ( for further details see [Equivalent circuit model](Equivalent circuit model)).

Furthermore, reducing the surface area of individual sensing pads to match the dimensions of individual cells enables an increasing of the density of electrodes on MEAs and of the spatial resolution.

The main advantage of using nanostructures does not lie, however, in the aspects mentioned above. In fact, the use on nanostructures permit to record action potentials in an attenuated patch clamp-like manner instead of a typical biphasic FP producing an intracellular recording. This technique allows high quality multisite, simultaneous, long-term, extracellular and intracellular recording and stimulation from many neurons under in vitro conditions. The development of intracellular recording and stimulation technologies enabled researchers to resolve subthreshold and synaptic potentials as well as to analyse the generation of action potentials. Moreover, current injection via these electrodes aims to repeatedly stimulate the neurons in addition to extract essential biophysical parameters such input resistance or membrane capacitance and examine synaptic properties.

The nanostructured electrode were firstly introduced by Spira et al in 2010 [43]. The recording of the intracellular AP was due to the neuron–electrode configuration used in his work, and allowed for the first time to record action potentials and synaptic potentials with a multielectrode array. In this paper, they show how the Aplysia neurons were forced to engulf the protruding mushroom-shaped microelectrodes (gMμE) with the activation of an endocytotic-like mechanism enabling both voltage recordings and application of current. The results were achieved with the generation of a high $R_{seal}$ and the increased junctional membrane conductance. It was shown that the recording mode could be switched from extracellular recording to intracellular-like recording by artificially increasing the seal resistance, introducing a layer of an RGD-based peptide surrounding the structures.

In the study from Park's laboratory [44], vertical nanowire electrode arrays (VNEAs) constituted of silicon dioxide with a doped silicon core tipped by Ti/Au were used to pierce the cellular membrane. This electrodes offer exceptional intracellular recording by penetrating the plasma membrane and directly accessing the cell interior while generating effective $R_{seal}$ with the plasma membrane. In the study, arrays of nanopillars were grown on the sensing pads. Embryonic rat cortical neurons or HEK293 cells were then cultured on the surface for a number of days. About 50% of the VNEAs spontaneously penetrated the plasma membrane of the embryonic rat cortical neurons or HEK293 as confirmed by the fact that current injection through the VNEAs generated a voltage drop across the plasma membrane. An electroporating pulse (approximately ±6 V, 100 ms) was applied to penetrate the membrane in the case in which spontaneous penetration of the membrane was not reached (for more details on the technique see Electroporation paragraph). Consistent with the intracellular positioning of the VNEA, all recorded APs were positive monophasic. Because of its small area, this nanostructure can address individual neurons by electrical stimulation and record the ensuing synaptic potentials thus recording APs from an individual neuron. However, at the same time, the coupling coefficient and signal-to-noise ratio are insufficient to enable recordings of subthreshold synaptic potentials because of the high impedance of the nanostructures.

Nevertheless, the application of current through the pillars, as was done in the previous-mentioned work, leads to the functional access in the intracellular medium. Different studies demonstrated that this localized membrane electroporation may lead to transient intracellular recordings of reduced APs. Xie et al. for the first time demonstrated electroporation of cultured cardiomyocytes by vertical nanopillar electrodes [45], Hai and Spira [46] demonstrated electroporation of cultured Aplysia neurons and Breaken et al. demonstrated single-cell cardiomyocytes electroporation using micrometre-sized TiN protruding electrodes [47].

In these studies, it was highlighted the low invasiveness of the method due to the fact that the intracellular access was transient suggesting that electroporation activates repair mechanisms that close off the nanopores. At the same time, because of this transient nature, the coupling between cells and electrode is not highly preserved during the whole recording. One way to solve this problem is that (as it was already demonstrated by the N. Melosh group [48] and applied by the C. Lieber laboratory [49]), a proper functionalization of the electrodes could facilitate the sealing formation reducing the attenuation of the recorded potential. Another way to solve the impedance problem could be increasing the density of nanopillar. Nevertheless, it was demonstrated that a too high number of them could produce the 'fakir' effect on the cell, which lay on the pillars tips without engulfing them. Bruggemann et al. fabricated densely packed vertical gold pillars pointing out that in this case the pillar nano-electrodes maintained an extracellular position [50]. Thus, the optimization of the pillar density and a correct functionalization could reduce the impedance allowing to solve the drawback related to this method.

Recently, another method, called plasmonic optoporation has been developed by using 3D plasmonic nanoelectrodes[51]. De Angelis et al. proposed this method in order to open transient nanopores into the cell membrane without compromising the seal between the cell membrane and the nanoelectrode, with no side effects for the cell. The process takes place just at the tip of the nanopillars when they are excited by short laser pulses, which produce hot-electrons and a nanoscale shockwave that can disrupt the cellular membrane [16]. Since the process is extremely localized, it enables a very stable intracellular coupling and a long-term observation[12]. Another advantage of the technique is that the optoporation is completely decoupled from electrical processes such as stimulation and recording. For all these reasons, it also does not perturb spontaneous cell activity and does not imply any recording blind time. The fabrication of plasmonic nanoelectrodes could be done on CMOS-based high density MEAs [10]and on flexible polymeric devices, which makes this technology appealing for high-density intra/extracellular recordings and integration with synthetic scaffolds either for *in vitro* or *in vivo* applications. This methodology demonstrates a vast potential and several appealing features to advance the quality of multisite electrophysiological recording technologies[52]. The technique presents several advantages but it requires, contrary to the simpler electroporation, a specific optical setup for functioning.

A brand new study that is important to be mentioned shows a novel type of nanopatterned volcano-shaped microelectrode that spontaneously fuses with the cell membrane and permits stable intracellular access. The complex ring-shaped nanostructures were manufactured following a simple and scalable fabrication process and provides passive intracellular access to neonatal rat cardiomyocytes [53]. Desbiolles et al. reported transmembrane action potentials with high spatial resolution without the need to apply physical triggers. This technique shows advantages for the assessment of electrophysiological characteristics of cardiomyocyte networks at the transmembrane voltage level over time.

Among the whole set of mentioned techniques, electroporation stands out for its potential application at large scale and as the most straightforward to be implemented in MEA devices. For these reasons and other (that will be discussed later on in this dissertation), we decided to exploit this technique within the development of this project.

For clarify and for better understand the technique, electroporation will be treated in detail in the next paragraph.

## 2.4 Electroporation

The phenomena of electroporation of the cellular membrane has been known for several decades, and it has recently received increasing attention for the manipulation of cells and tissues [54]. An electric field is usually created by applying an input voltage between two electrodes, thus, inducing nanometre-sized pores in the cell membrane. The mechanism of cell electroporation involves three steps: membrane charging, pore nucleation and evolution. During the first step, the non-electrically conductive cell membrane behaves as a capacitor between the conductive cell culture media and cytoplasm (for details on cellular membrane see cellular membrane composition). This electric charging of non-conducting lipid bilayers and omnipresent thermal fluctuations collaborate to create and enlarge a varied number of pores in the membrane itself. The subsequent formation of nanopores is triggered when the electric potential difference across the cell membrane rises. In fact, as membrane potential increases, hydrophobic defects (where the lipids are simply parted with respect to an intact membrane) of thermal origin rapidly expand into hydrophilic pores (toroidal pore stabilized by lipid head groups)[55][56]. The application of a transmembrane potential modifies the free energy of the hydrophilic pore such that its free energy. At a critical potential values, the defect could grows until the bilayer is destroyed.

Depending on the orientation of the electric field, these small pores may be unequally distributed across the cell surface. This bulk electroporation tends to create large numbers of small pores over a large fraction of the cell membrane because the whole membrane is subjected to the applied external electric field. The density, location and size of nanopores could be governed by the amplitude, duration and frequency of the input signal. The appropriate placement of the electrodes is crucial for the correct functioning, reproducibility and reliable incomes[57][4].

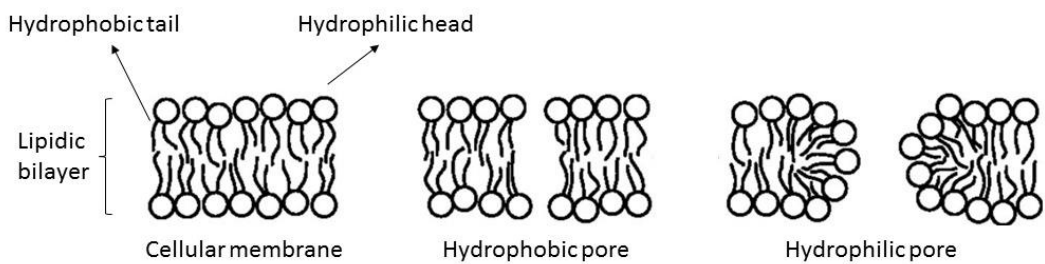

**Figure 5** *Representative schematic of the cellular membrane and hydrophobic and hydrophilic pores*

When an excessively high input voltage is applied, electroporation becomes irreversible, resulting in cell lysis. Bulk electroporation systems suffer from the need for high input voltage, non-uniformity of the electric field, formation of bubbles, and variations in local pH and temperature. These drawbacks can be overcome by miniaturization of the electrode in order to reduce the required input voltage, create a more uniform electric field, and rapidly dissipate heat because of the large surface-to-volume ratio. Applying this toward adhered cells has resulted in localized electroporation methods that have recently been developed in configurations suitable for single cells using fluidic nanoprobes and for multiple cells using a lab-on-a-chip approach (such as MEA devices).

In localized electroporation, the sharp and small shape (typically ranging from tens to hundreds of nanometers in diameter), together with the tight coupling, induce a larger electric field confined in a smaller area of the cell membrane. Owed to the focused electric field, this technique results in formation of relatively larger pores in a small area with the use of orders of magnitude lower voltages than typical bulk electroporation. The voltage applied at the electrode results in a transient and local increasing in the permeability of the cell membrane. These small pores permit to directly connect the interior of the cell with the electrode, enabling an intracellular recording emulating the patch clamp technique. In fact, this transient electroporation drastically improves the quality of signal recorded by the nanostructures by lowering the impedance between the electrode and the cell interior. The nanopillar electroporation causes minimal damage to the cell since a relatively low voltage is applied, and electroporation happens only in the membrane immediately surrounding each electrode. As a result, cell viability is higher than 90% for most cell types comparing with the bulk counterpart. Nanopillar intracellular recording is thus minimally invasive and also provides details of electrogenic cells action potentials with high resolution. After the appropriate electric pulse occurred, the nanopores shrink and reseal in a time range of seconds or minutes [58][59]. This specific feature consents to transitory enable and disable this way of access as needed.

Because the effective electric field applied to target cells depends on the microdevice configuration and dimensions, simulations of localized electroporation are often utilized to optimize and quantify the local electric field needed for the purpose[60].

# 3 Intracellular delivery

## 3.1 Significance

Injecting exogenous molecules into cells is a commanding strategy to treat and study diseases mechanisms, interpreting cell functions, reprogramming cell behaviours and investigate gene functionalities. The methods for delivery biomolecules such as DNA, mRNA and proteins into the cells are essential for cellular manipulation, genome engineering, cellular imaging, and medical applications [61][62][63][64][65]. In this regard, for example, recent developments in molecular biology and genome science have led to the discovery of a number of diseases-related genes. Attempts to apply these findings to the therapy of genetic and acquired diseases, including cancer and viral diseases (such as AIDS) are in progress. Gene delivery, antisense therapies and RNA interference could be considered promising therapeutic methodologies [66]. RNA delivery therapy is also identified as diagnostic markers and therapeutic targeting potential for many stressful and untreatable progressive neurodegenerative diseases such as, Alzheimer's, Parkinson's diseases and amyotrophic lateral sclerosis.

These treatments, together with many other pharmaceutical agents, including various large molecules (proteins, enzymes, antibodies) and even drug-loaded pharmaceutical nanocarriers, need to be delivered intracellularly to exert their therapeutic action inside the cytoplasm or onto the nucleus or in other specific organelles, such as lysosomes, mitochondria, or endoplasmic reticulum. In addition to enable the actual functioning of some treatments, it has been demonstrated that this intracellular delivery of various bio-compounds, DNA, RNA, plasmids and drug carriers can dramatically increase the efficiency of various treatment protocols.

Despite this great importance, the efficiency with which these treatments are delivered to cells remains low. The focal problem of therapeutic effectiveness lies in the crossing of cellular membranes. The cell membrane, in fact, prevents big molecules, such as peptides, proteins, and DNA, from spontaneously entering cells unless there is an active transport mechanism. Significant effort has been made, during this century, to develop agents which can cross the cellular barriers and deliver therapeutic agents into cellular compartments.

## 3.2 Cellular membrane composition

Biological membranes are supramolecular assemblies composed of a lipid double layer interspersed with transmembrane proteins. Each monolayer of the membrane consists of several adjacent lipid molecules, which are composed of typically two acyl chains, called hydrophobic tails and a hydrophilic headgroup. The two monolayers, facing each other with the hydrophobic tails, function as a barrier of 3 – 5 nm thickness, is not soluble behaviour in aqueous environments [67]. Membrane lipids are principally of phospholipids and sterols (usually cholesterol). The membrane separates the cytoplasm from the extracellular environment and because of its lipidic nature exhibit partial permeability to some small hydrophobic and polar molecules. Lipid-soluble molecules and some small molecules can permeate the membrane, but the lipid bilayer effectively repels most of the water-soluble molecules and electrically charged ions that surrounds it. The interspersed proteins serves as channels for the exchange of ions. The two main types are ion channels and ion pumps. The first ones, are proteins endowed with a pore that crosses them in the centre, and allow a determined type of ions to flow inside them along the concentration gradient. Ion pumps instead use energy for the active transport of ions through the membrane against their concentration gradient.

Particles too large to be diffused or pumped are often swallowed whole by an opening and closing of the membrane. The cell membrane itself undergoes intensive movements in which portion of the fluid medium outside of the cell is internalized (process called endocytosis) or portion of the internal medium of the cell is externalized (exocytosis). These activities involve a fusion between membrane surfaces, followed by the re-formation of intact membranes.

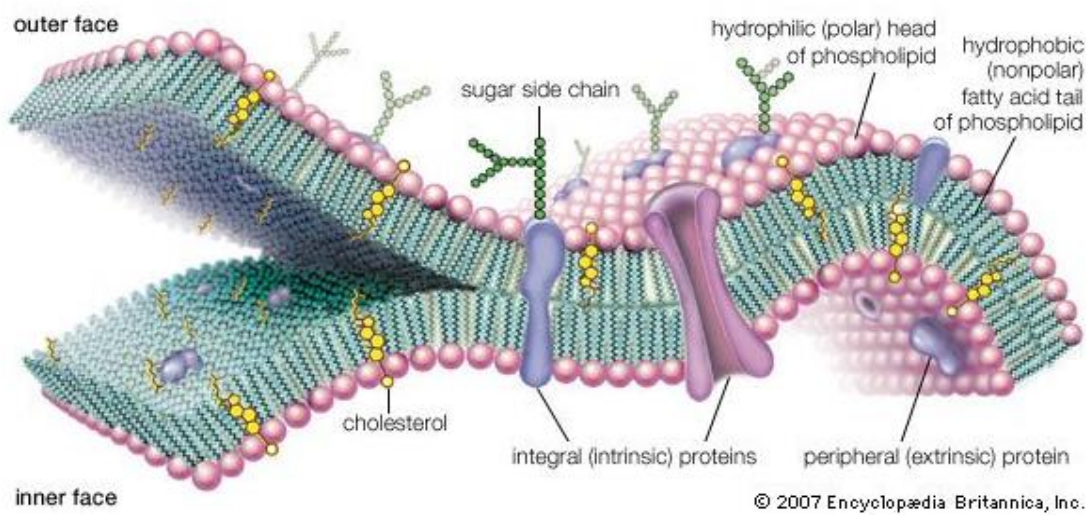

**Figure 6** Cellular membrane. *Encyclopedia Britannica*

## 3.3 The development of the technique

The field of intracellular delivery could be dated back to the advent of membrane-microinjection in 1911 [68]. The method described there makes possible for the first time the segregation of one or more micro-organisms with the simultaneous injection of them into the cytoplasm of living cells. In order to accomplish this, pipettes were finally constructed as to minimize the injury to the injected cells and at the same time having sufficient rigidity to pierce the cell wall. An injection force was employed in order to overcome cell pressure, capillarity, and any obstruction in the pipette. From that moment on, a broad range of options have evolved. Nowadays, we can classify the developed methodologies into two classes: carrier-mediated and membrane-disruption-mediated intracellular delivery.

For carrier-mediated approaches, the first research involved the use of several cationic compounds merged with the negatively charged nucleic acids so as to facilitate the uptake of DNA and RNA. Examples include precipitates formed with diethylaminoethyl (DEAE)-dextran and calcium phosphate [69]. Inspired by these initial findings, chemical complexes and modified viruses were subsequently deployed as tools for DNA transfection.

In membrane-disruption-based approaches, the initial attempts in the field regarded low-throughput microinjection [70] and membrane perturbation [71]. The demonstration of DNA transfection by electroporation in 1982 [72] ignited a wave of experimentation and development of new approaches of membrane-disruption methods.

## 3.4 Methods

### 3.4.1 Carrier based

The carrier-based approaches include several biochemical assemblies mostly from molecular to nanoscale dimensions. The carrier has the labour to package the cargo and protect it from degradation in order to gain access to the intended intracellular compartment and to release the load with the proper dynamics. Carriers can be bio-inspired, such as re-formed viruses, vesicles and functional ligands and peptides. Alternatively, they may be based upon synthesis techniques involving assembly of nanoparticles and macromolecular complexes from organic to inorganic origins. Most carriers cross the cell membrane via endocytosis but some may display the ability to merge directly with the target cell membrane. In the case of transfection (nucleic acid delivery), vectors are designed as constructs able to contain foreign DNA for expression or replication inside the new cell [73]. The principal vector types are plasmids, cosmids, artificial chromosomes and viral vectors. Viral vectors alone are capable of unassisted entry and at present, are the most clinically advanced nucleic acid delivery agents thanks to their high efficiency and specificity. These tools have been implemented in clinical trials for decades, being considered as the most promising approach for gene therapy [74][75]. However, viral vectors present drawbacks such as immune response, complexity of preparation and safety and so viral systems have struggled to obtain FDA approval.
Motivated by these limitations, hundreds of non-viral vectors and synthetic carriers have been designed, using massive combinations of lipid, inorganic materials and polymers. These materials are sometimes functionalized with ligands, cell-penetrating peptides and other targeting agents in order to be uptake by the cell. Most of these new carriers are designed for nucleic acid transfection, but recent efforts pursue to expand their ability to deliver proteins or other biomolecules. Mostly, they are up-taken by specific endocytic mechanism based on cell surface interactions and physicochemical properties. The most advanced non-viral vectors for nucleic acid are considered the lipid nanocarriers. The problem of most of the previous mentioned methods, however, is that any object entering the cell via the endocytic pathway will be captured in endosome and eventually will end in the lysosome, where active degradation processes under the action of the lysosomal enzymes occur. Consequently, only a small portion of unaffected substance will be effectively delivered in the cytoplasm. Moreover, toxicity of carrier material and distortion in the membrane transferring processes have been noted in multiple cases [76].

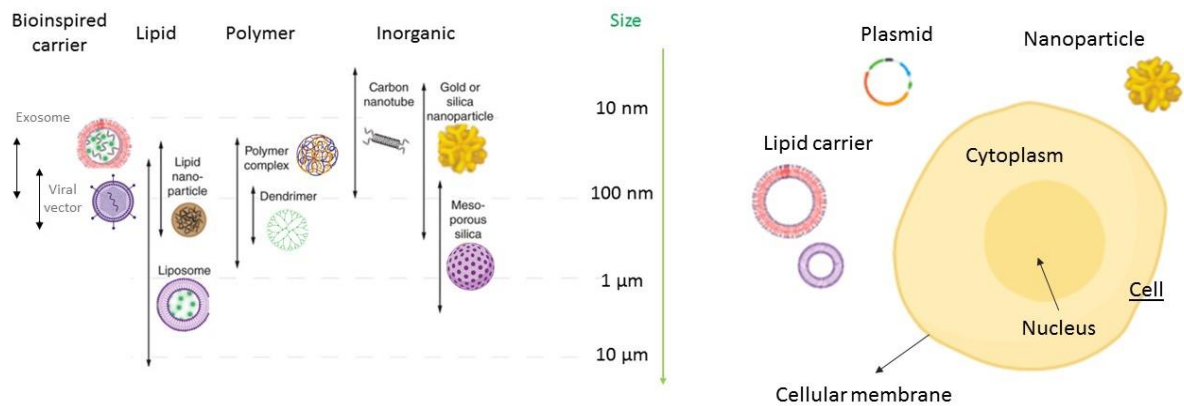

**Figure 7** *(left) Carrier mediated intracellular delivery architectures with approximate size range and (right) schematic of a cell and the mentioned methods*

Carriers with fusion capabilities evade endocytosis by releasing their cargo directly into the cytoplasm. More recently, cell-derived vesicles known as exosomes have been demonstrated to fuse with target cell membranes for the exchange of RNA and proteins between immune cells. Even though the exact fusion mechanisms needs to be clarified, these bioinspired systems may represent a new cohort of carriers with which to overcome the poor efficiency and toxicity of the synthetic counterparts [77].

The major restraint of carrier systems is the limited combination of possible cargo materials and cell types. Target cells may not display the appropriate surface interactions, receptors, endocytic activity, or endosomal escape pathways. Furthermore, the properties of the cargo materials such as, hydrophobicity, size, mechanical properties, composition and charge often make the behaviour unpredictable. They may not efficiently composite with the carrier, tolerate to be packed, unpacked properly, or may not be able to deliver sufficient quantities for a given application. All these carrier mediated intracellular delivery methods present then limitations due to an incomplete comprehension of their reaction with biological environments, highly stochastic dosage, lack of specificity, low efficiency, high expenses and the toxicity issues related to these cutting-edge materials.

### *3.4.2 Membrane disruption*

Membrane-disruption modalities are mainly physical, involving the introduction of transient discontinuities in the plasma membrane by mechanical forces, electromagnetic radiations, thermal deviations, or by appropriate biochemical agents, such as detergents and pore-creating toxins.

The first method to be implemented and the most straightforward is the mechanical disruption, where tensile strains rupture the lipid bilayer of the membrane. Such mechanical poration can be obtained by solid contact [78], osmotic pressure [71] or fluid shear [79]. These methodologies present different disrupting mechanisms that depend on the contact area and strain rate. The applied force may both disrupt the membrane immediately, such as in the case of sharp objects like microneedles (they concentrate the force to a small region and penetrate rapidly) or first deplete membrane reservoirs such as in the case of osmotic shock.

Thermal deviations may promote membrane defects through numerous mechanisms. The first consists in a more intense molecular fluctuations and subsequent dissociation of lipids in the membrane due to the higher kinetic energy associated with higher temperatures [80]. Second, inside the physiological range (0–40 °C), inducing rapid thermal phase transitions may lead to the generation of holes. The last, instead, at temperature close to the 0°C, the formation of ice crystals can activate mechanical expansion and cracking of the cell membrane, which may be reparable upon thawing [81].

Another important method to be named is electroporation. The physics of this method is widely described in the Electroporation chapter. Electroporation rapidly gained a base as commercial products since when it was launched from the mid-1980s. Most other membrane-disruption techniques were not broadly adopted, presumably owing to skill-dependent operation, need for specialized equipment, high cost, higher invasiveness or limited throughput. Electroporation presents the capability of opening transient pores into the cell membrane in order to access the intracellular environment allowing for a time controlled and a rapid delivery [82] [83]. The system classically consists of metallic electrodes, a pump for fluidic control, and either a fluidic pipette or cantilever with nanoscale tip or a microporous substrate combined with a microfluidic device. Localized electroporation shares the same physics as bulk electroporation, but the concentrated higher electric field and nanochannel shape, size, and interface allow a much better control of the effects on the cells. This extremely large electric field, created just in the immediate proximity of the cell membrane, is orders of magnitude larger than typical inputs for electrical cell lysis techniques and results in the formation of large pores and a strong electrophoretic force that can transport large molecules directly into the cytoplasm [84]. The technique is considered one of the most promising tools as it is not dependent on carrier or cell properties, granting to inject any material dispersed in cellular medium and applicable to a broader range of cell types.

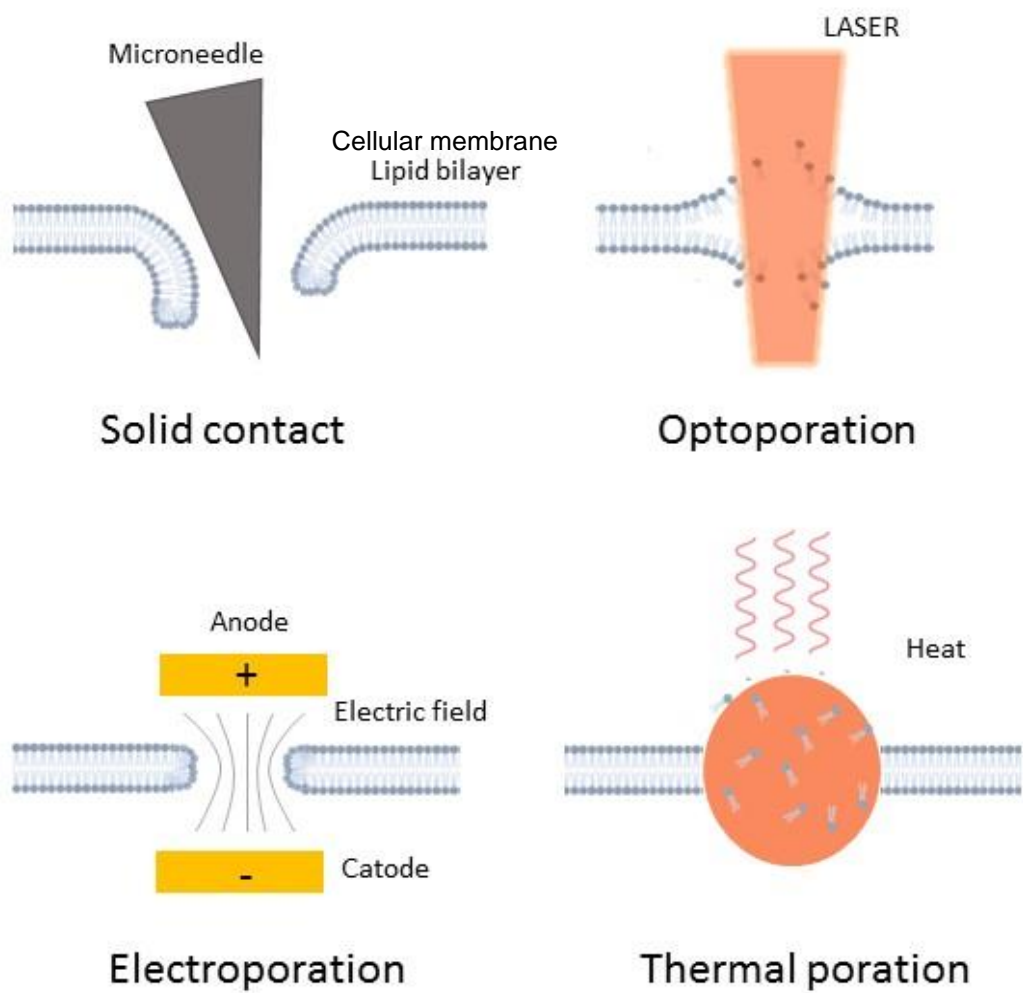

**Figure 8** *Representative schematic of the mentioned physical methods for intracellular delivery*

A valid alternative is optoporation, which exploits laser sources to produce holes at a localized site on the cell surface, being less invasive to the cell membrane. This process may involve mechanical, thermal and chemical effects, and, depending on the laser parameters, could bring to a really low invasiveness [85][86]. Despite this, a strong drawback that prevents the commercialization is the low throughput of this technique and the need for a specialized setup for combining the laser source and cell cultures.

All the above mentioned methods are less dependent on cargo properties comparing with the carrier-based methods, being able to deliver almost any material dispersed in solution. These physical methods, allow to have a feasible control and efficiency in cell transfection and desirable delivery for a huge variety of cargos, from small molecules to larger proteins or antibodies [86]. Moreover, the ability to turn ON and OFF quickly membrane porations effects enables temporal control and rapid delivery. A further strength of membrane-disruption techniques is the broad range of cell types that can be addressed.
Traditionally, the major weaknesses of membrane-disruption strategies have been the excessive cell damage caused by bulk methods, the poor scalability (mostly in mechanical poration) and the poor understanding of cell recovery mechanism. However, membrane disruption has seen promising progress in recent years through nanotechnology, microfluidics, and laboratory-on-chip devices. Recent advancements in nanofabrication techniques have enabled the realization of nanostructured platforms for improved intracellular delivery, providing selectivity, high spatial and temporal resolution of the delivery process.

# 3.5 On chip devices for intracellular delivery

The traditional transfection methods described above usually require cell suspension, which may disturb cellular pathways under investigation and are often extremely harsh for sensitive primary cells. These disadvantages are particularly problematic for in vitro studying where cells adhere to the device. Primary cells such as neurons need to be subjected to transfection mechanism in order to explore the pathogenic mechanisms of neural diseases and to develop gene therapies for disorders such as Alzheimer's, Parkinson's, epilepsy, and many others.

Several technologies for studying cultured cells have been developed in the last decades, among all micro and nanotechnologies offer individual cell access and non-destructive probing. Nanotechnologies provide unprecedented levels of spatiotemporal control and highly minimize cell stress. These factors enable high viability delivery of biomolecules, high-efficiency and in some cases non-destructive cell analyses that could be determinant for exploring time-dependent differentiation mechanisms, phenotypes and heterogeneity. These technologies have proven encouraging potential as alternative methods for molecular delivery into cultured cells utilizing working principles that include mechanical penetration and localized electroporation.

Exploiting these capabilities, electroporation technologies display promising potential for temporal studies of gene expression and cell phenotype of *in vitro* cells that can provide information in systems-biology analyses, for example during stem cell differentiation. In this regard, integration of biosensor with different functions, such as transfection, intracellular sampling, and biomolecule detection with single-cell specificity and high throughput, would provide powerful tools for advancing applications in therapeutics, diagnostics, and drug discovery, particularly for cellular engineering involving cell reprogramming, stem cell differentiation, and gene editing. In this direction, nanostructures such as nanopillar [6], nanostraws [5] and nanofountain probe [7] where developed in the last years by different groups in order to reach an efficient and localized delivery on a chip. The nanostraws device proposed by Melosh group demonstrates to achieve high efficiency molecular delivery and high transfection into mammalian cells. The platform consisting in arrays of alumina nanostraws above microfluidic channel trough non-destructive nanoelectroporation yields the delivery of such molecules with high spatial resolution [87]. In 2013, Espinosa demonstrated the electroporation performed by his nanofountain probe on single HeLa cell within a population. They were able to transfect these cells with fluorescently labeled dextran and imaging the cells to evaluate the transfection efficiency and cell viability [7]. One year later, Ying Wang et al proposed a novel platform for delivering materials into living cells using mechanical disruption of the membrane obtained by an array of diamond based nanoneedles [6]. The device was used to deliver genetic materials, especially plasmids, into neurons. In comparison with the previous presented device, they were able to reach a large population of neurons but could not control the delivery up to single cell resolution.

Despite the efficiency or high specificity of the intracellular delivery achieved thanks to the nanoscale shape of these electrodes, these devices lack of a proper integrated system for monitoring cells behaviour during their development or growth in *in vitro* condition and evaluate the effect of the molecules injected on the cell culture.

One way to explore the complex behaviour of stem cells, or neuronal networks and cardiac cells, could be to integrate the capabilities of microelectrodes devices (that include long-term cell culturing, non-destructive cell accessing and high quality recording capabilities), with high spatio-temporal resolution drug delivery. More specifically, in order to comprehend complex intracellular interactions and to develop models for cellular behavior, it is essential to combine in the same device tools for cell stimulation and simultaneous probing. These combined systems will have a significant impact on fundamental biological studies and will lead to improvements in our capability to understand cell behaviour and to develop predictive analyses for engineering systems such as tissues and therapeutics. My project fits in this context, so as to cover this demand.

# 4 Microfluidic multielectrode array for intracellular drug delivery and recording

## 4.1 Microfluidic MEA concept

The goal of this project was to develop a novel microfluidic device to optimally culture cells during expansion, efficiently deliver molecules into these adherent cells by localized electroporation and record their spontaneous electrical activity during these processes.

As pointed out in the previous chapter ( Multielectrode arrays), complex processes such as neurological systems have important biochemical components in addition to the electrical components of signalling. Studying the function and connectivity of neurons requires a comprehensive study of both electrical and chemical aspects of the cell culture. In vitro screening platforms should recreate an environment that is as representative as possible to the one found in vivo. Physiologic, pathologic and pharmacologic responses could be investigated with a precise control of the environment surrounding by controlling the reagents and factors distributed to these cells via the microfluidic channels to spatially match the location of the cells in order to study the effect of the treatment on just smaller parts of cell culture.

For this purpose, the multiple capabilities of the hollow nanostructures, developed in the previous years by the group [88], have been exploited in such a way to create the abovementioned platform. The hollow nanoantennas, were used as a hybrid structure for a high performance intra and extracellular recording [51] and as a nanoscale tool for selective and intracellular delivery [86].

The used approach resumes the well-established passive MEA concept enriching it with the possibility of cell poration, and of localized and carrier-free drug delivery. As depicted in the picture below (Figure 9), the Microfluidic MEA (MF-MEA) is composed by a thin silicon nitride layer on top of bulk silicon and two microfluidic channels that runs underneath the surface. On top of these two compartments, on the other side of the device (top side), we can find 24 planar gold electrodes precisely distributed. The nanostructures fabrication process (explained in detail in the next paragraphs) allows for simultaneous shaping of the nanostructures situated in correspondence of each planar electrode and the piercing of the thin surface over the microfluidic channels (Fig 10). Exactly this protocol allows the creation of the flow-through design of the device. More in details, the molecules injected in the two chambers in bottom side of the device could reach the top side through the hollow nanoantennas and reach the cells cultured on top of them. The two microfluidics compartment enable the concurrent delivery of two different molecules on the culture without cross contamination.

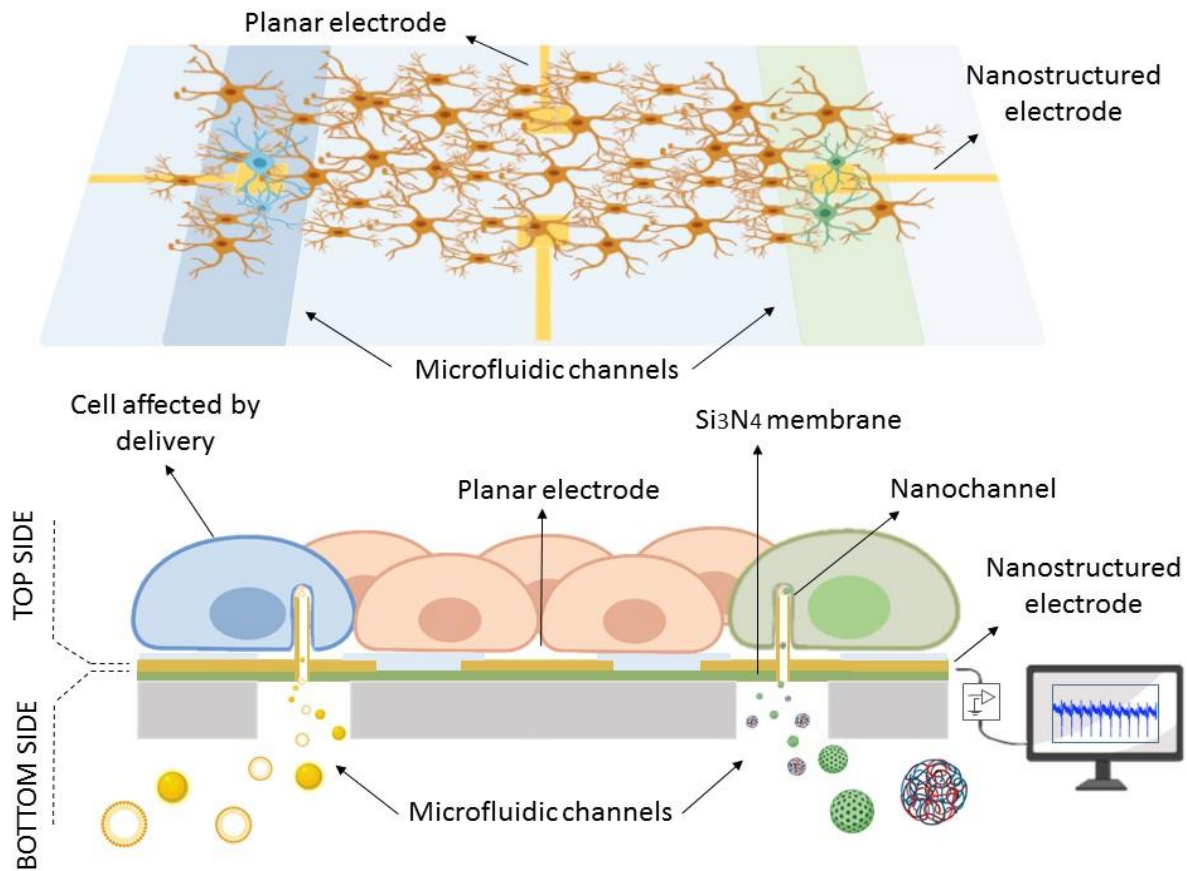

**Figure 9 Microfluidic multielectrode array biosensor idea.** *Microfluidic channels on the bottom side of the device keep the exogenous molecules separate from the cellular medium until they are spread by the nanopillars. The hybrid nanostructures improve the extracellular recording, enable the intracellular monitoring while allow high spatially localized delivery*

These structures permit to combine microfluidics with electroporation performed by the nanoelectrodes, allowing for a precise control of the biomolecules delivery. The nanopillar shape, moreover, permits to perform high quality extracellular recording and to switch to intracellular recording applying the same electroporation protocol.

The device so constituted, can allow for monitoring progressive changes in electrophysiological behaviour of the cell under examination at different sites simultaneously with the high accuracy achieved through the increased coupling between the cell and the 3D nanostructures. In this way, the device configuration could provide versatility in controlling the biochemical elements that influence cells activity and growth, could enable the formation of biochemical gradients, the injection of specific treatment or molecules and at the same time allows for a real time monitoring evaluation and analysis. The device is enriched with the capability of enducing pathology or treatment overpassing the cell membrane with almost single cell resolution. This approach could be used in cancer therapeutics research, growth factors studies on neuronal stem cells differentiation or simply, because of the complex interactions that occur among neural lineage cells, in vitro studies of neuronal networks.

The performances of the device will be examined evaluating initially the cell viability on the device. Because of the presence of the microfluidic channels, the response in cells growth to the non-uniformity of the surface has to be delaminated. To reach this goal, both cardiac and neuronal cells will be cultured on the platform. As second step, the intracellular delivery mechanism will be inspected with a simple fluorescent molecule in order to clearly see the effect. At this point, electrical recording could be performed on electrogenic cells and the delivery of more interesting molecule could be completed.

## 4.2 Results and methods

### 4.2.1 Biosensor fabrication

#### 4.2.1.1 Microfluidic channels and membranes

In order to reach the flow-through design, the typical elements of the MEA structure (electrodes, feed lines, and passivation layers) are fabricated on a 525 µm thick silicon wafer with 500 nm $Si_3N_4$ layers on both sides.

Starting from one side of the substrate, the $Si_3N_4$ and the Si are selectively removed trough dry and wet etching, forming two thin nitride membranes (Fig 10). More in detail, a 400 nm Cr mask was sputter coated on this side and a thin layer of photoresist was spin coated on top of it. UV lithography allows to expose the desired area and after 60s of development (MF-319 microposit developer) we could remove the Cr at the membranes and channels area via chemical etching. The remaining resist is then dissolved in acetone, leaving the Cr as a mask for reactive ion etching.

A mixture of gases, $CHF_3$ and $O_2$ (tuned at 70 and 10 SCCM, respectively) were used to selectively etch the 500 nm thick silicon nitride. The pressure, temperature and power were fixed at 1 Pa, 20 °C and 30 W, respectively. As a last step, the Cr mask was chemically removed and Si was wet etched in a solution of 1:2 potassium hydroxide (KOH) in deionized water (DI).

Given the selectivity of the KOH etching, the process stops when the nitride is reached at the other side of the wafer. The final depth of the channels can be varied accordingly to the geometrical parameters of the etching area (Fig 10 B and D). This specific microfluidic design can be easily changed from 2 to 6 independent channel configurations (Fig 10 C), for allowing to inject more types of biomolecules on the same device or perform multiple analysis with the same platform.

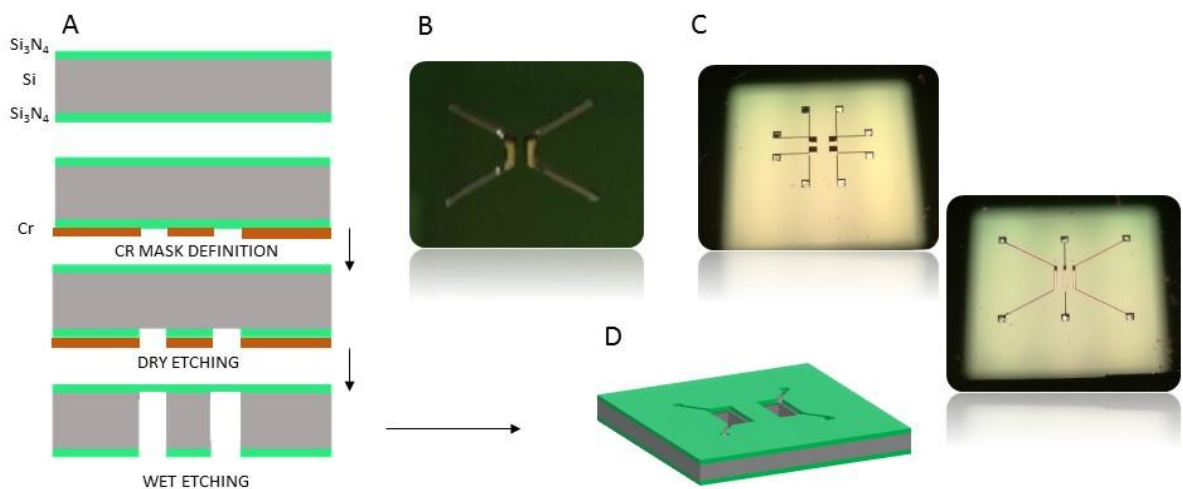

**Figure 10** *A) Microfabrication step for manufacturing channels and membranes from the top to bottom. D) sketch of the final step B) Picture of the fabricated membranes and channel on the bottom side of the device C) Pictures of the fabricated devices with higher number of membranes and channels.*

## *4.2.1.2 Electrodes fabrication*

The MEA enriched with 3D nano-electrodes were fabricated using the well known technologies developed in other works [89][88] and briefly explained in the first chapter. In this process, a thin layer (≈2 μm thick) of S1813 was spin-coated on the silicon nitride membrane, on the planar side of the substrate. After UV exposure and 60 seconds of development in MF-319, 24 Ti/Au electrodes, conductive tracks and electrodes were fabricated through electron beam evaporation in a high vacuum chamber (base pressure $10^{-7}$ mbar) with a 0.3 Å s$^{-1}$ deposition rate. The unexposed resist and the metal on top of it were removed through a conventional lift-off process in Remover PG (MicroChem) at 110°C. The substrate was then rinsed out in isopropanol (IPA), and the residual photoresist is ashed away with $O_2$ plasma at 100 W for 5 minutes. At this point the 24 electrodes of varying dimensions (from 100 to 900 μm$^2$) are defined using a passivation layer of SU-8 that covers the device leaving openings on the electrode sites. The MEA, therefore, presents 24 electrodes placed in a 4 × 6 matrix and with an inter-electrode distance of 400 μm, resulting in an active area of 2 × 1.2 mm$^2$ (see fig 11A).

The gold coated 3D nanoelectrodes were fabricated on the planar electrodes by focused-ion-beam (FIB) lithography (with the exposure dose at 27 nC μm$^{-2}$) from the backside of the $Si_3N_4$ membrane, using the well known protocol previously developed [88]. As it is possible to notice from the picture below (Fig 11A), the obtained hollow nanotubes are characterized by a slightly conical shape, with an external diameter of about 300 nm and a height of 1.3 μm. The so fabricated hollow structures are afterwards sputter coated with a 40 nm gold layer to make them conductive and biocompatible. Because of the fabrication method, the gold nanoelectrode arrays are realized only on those electrodes lying on the nitride membranes. The number of nanostructures per electrode depends on the desired microfluidic configuration. In our application we choose to fabricate 16 nanoantennas on each planar electrode.

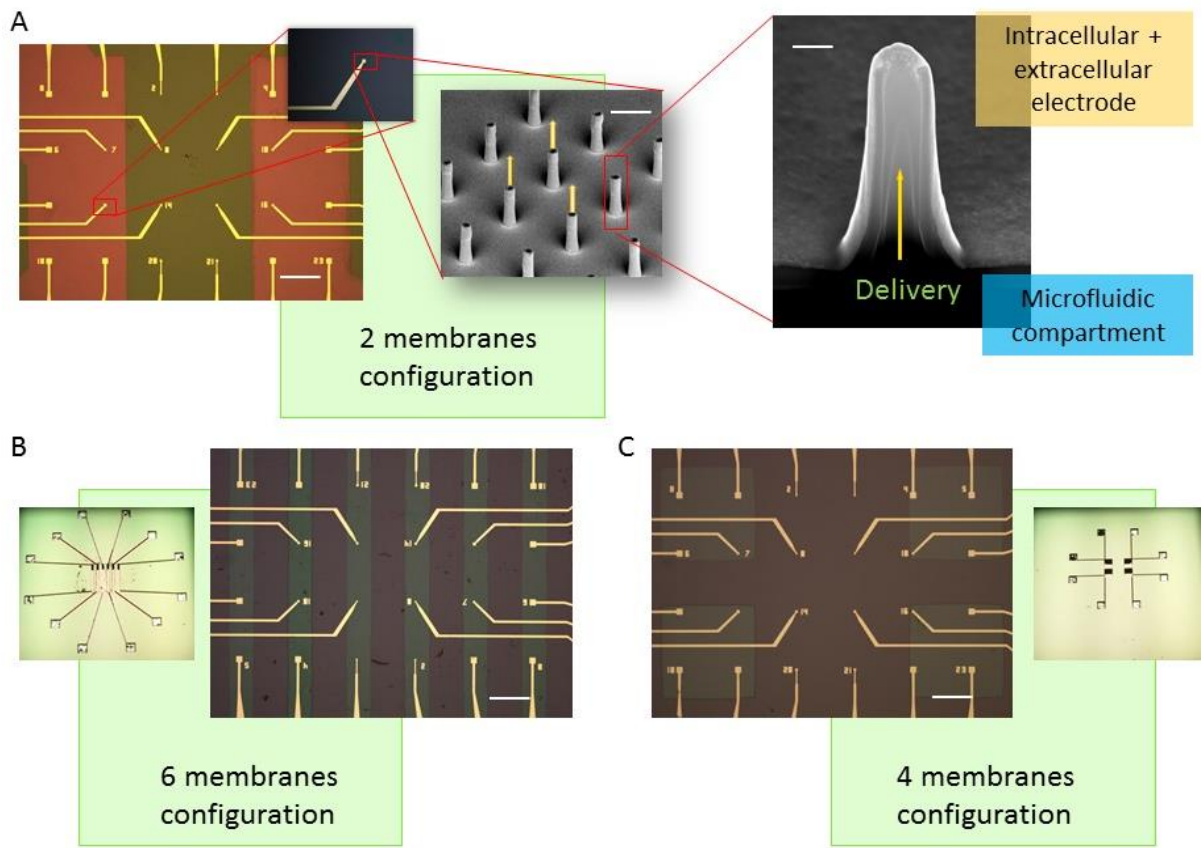

**Figure 11** *A) Bright field image of the 24 electrodes on top side of the microfluidic channels (pink) and on the bulk surface (dark green) Scale bar: 300 µm. On the right a SEM image of the 3D hollow nanoantennas on the electrodes (scale bar : 2 µm) and a cross section of one of those (scale bar: 200 nm). B) e C) Top side of the device in the case of 4 or 6 membranes respectively.*

The specific geometric characteristics of 3D hollow metallic nanoelectrodes, as already pointed out in the previous section, allows to apply electroporation protocols to the cells and to get direct access into the cell interior. These hybrid structures, in fact, on one side act as nano-fluidic channels for the flow of molecules or drugs from the underlying channels, and on the other side offer an improved electrical recording, increasing the cell-electrode seal resistance and thus reducing parasitic signal losses. This design provides a way to deliver molecules locally only to the cells lying on the nanotubes and a high quality cellular recording on a large population of cells.

## 4.2.1.3 Final steps and PCB interfacing

The so constituted device with nanostructured electrodes was mounted on a PCB (printed-circuit-board) and passivated with a 2 mm thick layer of polydimethylsiloxane (PDMS), leaving a 4x4 mm open window in the centre of the device. This aperture confines the cell growth area at the active area of the device, while the PDMS passivation layer covers and insulates the electrical connections with the PCB ensuring non-toxicity, inertness and biocompatibility. A glass ring was attached to the MEA by means of PDMS to preserve the cellular medium in the same way of a culture well (see picture below). On the bottom side, the open micro-channels were plasma bonded with a piece of polymerized PDMS. Tight sealing between $Si_3N_4$ and the PDMS was guaranteed by chemical activation of both surfaces via oxygen plasma treatment for 30 s at and 30 W. The PDMS bonding step assures the realization of two independent chambers with inlets and outlets, enabling the selective delivery of reagents in different areas of the MEA. As a last step, tubing was added to provide external connections to the pumping system. Even with the integrated multiple features, the platform still remains compatible with standard MEA acquisition systems.

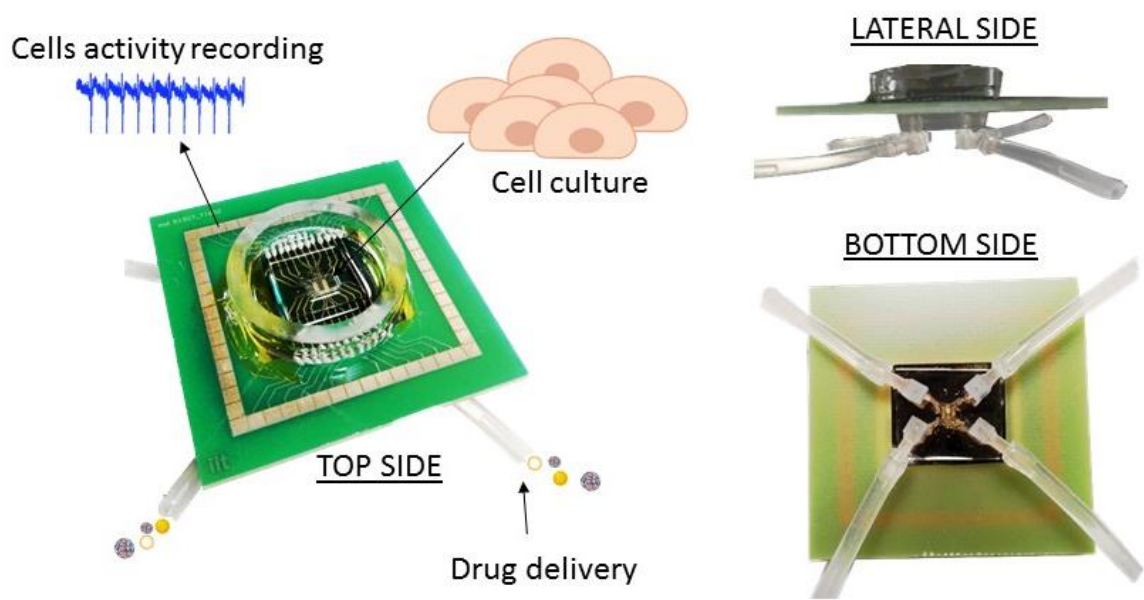

**Figure 12** *Picture of the top, lateral and bottom side of the final device*

This approach has the advantage that the molecules flow beneath the cell culture and are directly injected into the cells through the gold nanotubes avoiding to be dispersed in the cell medium, which remains mostly uncontaminated. The drugs could be easily washed away into the tubing without affecting the cells on the top side of the device. This makes feasible a subsequent injection of different drugs on the same target culture.

## 4.3 Neural and cardiac cells viability

### *4.3.1.1 Cell viability in HL-1 cells culture*

In order to investigate the effects of culturing the cells over a heterogeneous sample consisting of areas on bulky silicon and other regions on the thin nitride membrane, viability tests have been performed.

As a first step, the micro-channels and tubing at the bottom of the device were manually filled with culture medium (Claycomb), in order to saturate the PDMS porous matrix. After 24 hours, the medium was completely removed from the culture well and HL-1 cells derived from rat cardiomyocytes, were seeded at a density of 35 000 cells per $cm^2$ and grown with the same culture medium. After 4 DIV the whole cultured cells were treated with DAPI (a fluorescence blue staining of the nuclei) injected directly from the glass culture well. The PBS was injected in the tubing underneath, so as to evaluate possible divergences in cell proliferation with some fluids in the nanostructures. The images (Fig. 13) reveal a homogeneous cell growth over the whole MEA surface, with no distinction between the nitride membranes and the bulk silicon. To further assess this element, the cell density was calculated for both bulk silicon and nitride membranes respectively and the values are $(21 \pm 3) \cdot 10^{5}$ cells/$cm^2$ and $(22 \pm 1) \cdot 10^{5}$ cells/$cm^2$. This guarantees that the device configuration does not influence the cell culture growth and development.

The figure below shows the upright images of the cardiomyocyte on both surfaces.

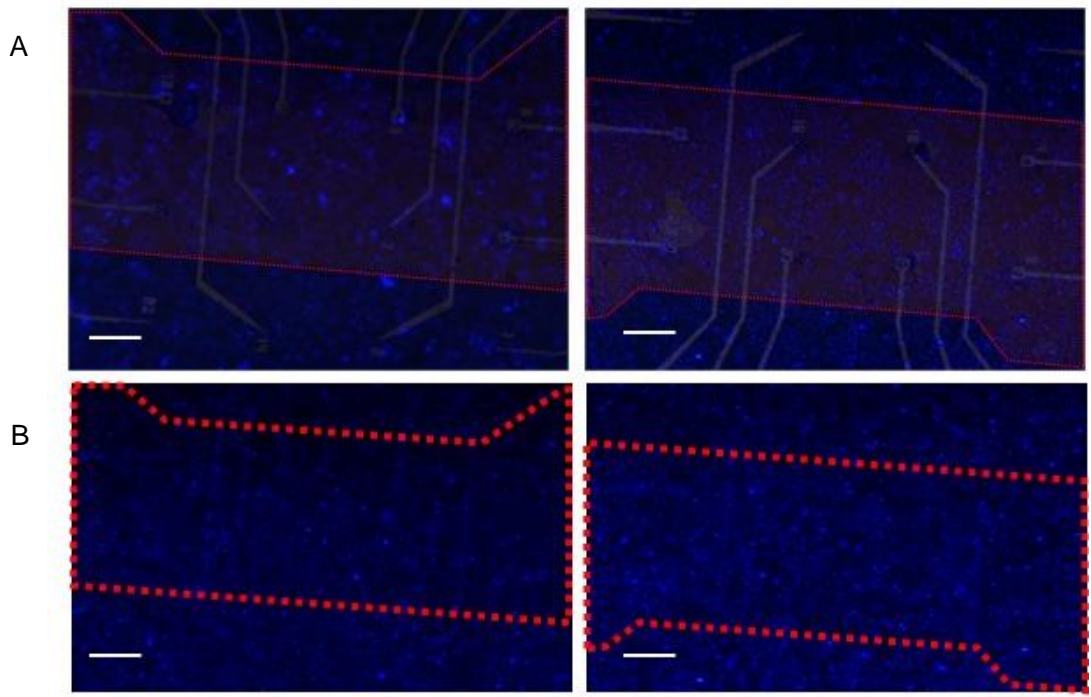

**Figure 13** *20 X Upright confocal microscope images of HL1 cells on the device. Scale bar 150 μm. The red dashed zone represent where was performed the counting of cells for both B) bulk silicon and A) nitride membranes.*

*4.3.1.2 Cell viability of hippocampal primary neurons*

MF-MEAs have been sterilized with 20 minutes-UV exposure in laminar flow hood for each side of the chip. Devices were pre-conditioned 2-days before cell seeding by incubation overnight at 37°C, in Primary Neural Growth Medium (PNGM). After 24 hours, devices were immediately coated with a solution of 30 μg/mL poly-D-lysine and 2 μg/mL laminin in PBS in order to enhance primary neuronal cells adhesion and proliferation on the devices and incubated overnight at 37°C. After 4 hours the substrates have been washed extensively four times with sterile water and let dry in sterile condition until the cell seeding. Hippocampal neurons were cultured at a density of $10^3$ neurons/mm$^2$ on top of the MF-MEA biosensor. Cultures were maintained for more than 3 weeks, while one-third of the medium was regularly changed every four days. To investigate the cell viability and the development of neuronal networks on the platform, confocal images of the neurons on the surface were obtained after fixation and staining. The immunofluorescence protocol was applied to the cells at 18 DIV, at the stage of a mature network.

The major microtubule associated protein is visualized with MAP2 marker (in green), astrocites marker GFAP are shown in yellow while the nuclei were stained with nuclear marker DAPI (in blue). As already mentioned for the cardiac cells, the comparison between the bulk surface and the thin surface with the microfluidic channels underneath is important in order to state the possible effects of a heterogeneous surface on the neuronal culture.

In the picture below (Fig. 14) are shown 60x confocal images acquired using a Leica upright microscope. It is possible to observe a detailed view on electrodes with and without the hollow nanoantennas that allow us to evaluate the viability of the cells on both nanostructured electrodes on the silicon nitride membranes and the planar electrodes on the bulk surface.

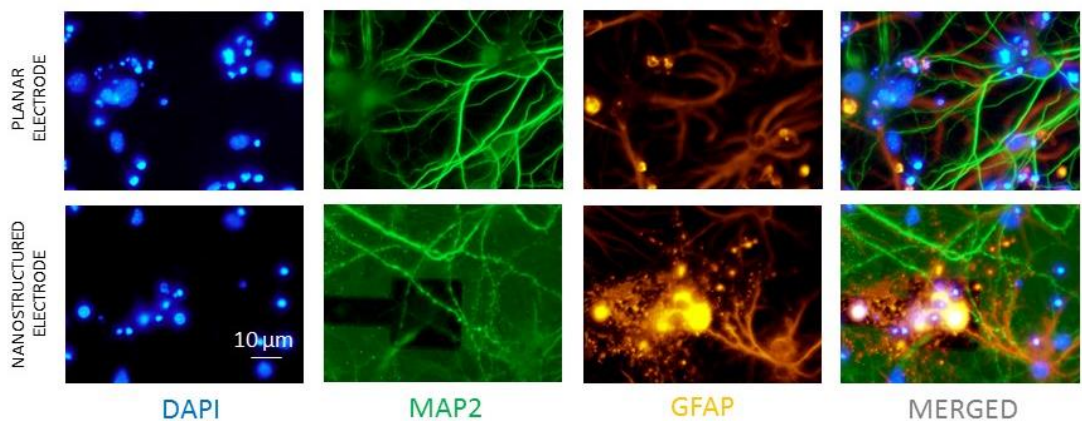

**Figure 14** *60X confocal images of neurons on both bulk surface (planar electrodes) and silicon nitride membrane (nanostructured electrodes)*

The MAP-2 expression shows a healthy and homogeneous neurite growth on both surfaces with no distinction. This heterogeneous device ensures in both cases a good network development demonstrating that the platform does not affect the cell culture attachment, the network development and processes formation.

# 4.4 Intracellular delivery in the HL-1 cellular model and simultaneous recording

In order to verify the success of our multifunctional platform on intracellular and localized delivery, we inject molecules into 4 different cultures through the 3D nanoelectrodes. The hollow nanostructure design enables to address just the few selected cells on the electrodes, without affecting the rest of the network. For this purpose, cardiomyocyte from HL-1 cell line were cultured on the active area of the device. The HL-1 cell line, having been characterized broadly in literature, represents a well-established benchmark model for testing the recording performance of the brand new MF-MEA sensors[90].

As a first step, when the cells reached the confluence, calcein-AM was delivered at a concentration of 200 µM in phosphate buffer solution (PBS) to the cells through the 3D nanoelectrodes from the microfluidic channels. Calcein-AM is a membrane permeant molecule producing green fluorescence at $\lambda = 515$ nm. The solution was at first incubated at 37 °C, and then was injected with a pumping system into the two isolated microchannels, allowing the fluorophores to diffuse through the nanoelectrodes. According to the tubing volume and design, a volume of almost 150 µl was required to fill the microfluidic channels underneath the MEA and the bottom part of the silicon nitride membranes.

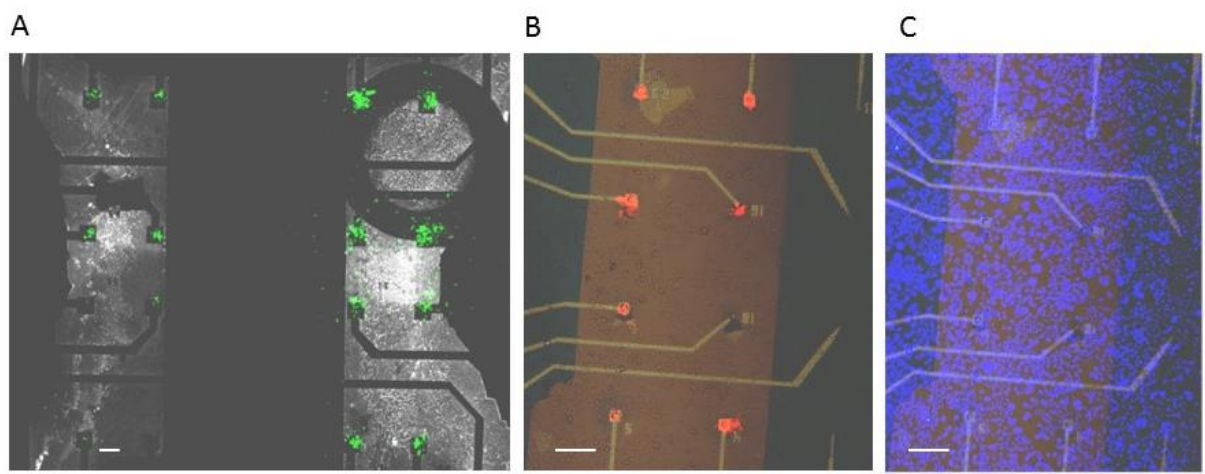

**Figure 15** *A) 5X fluorescence microscope image to the MF MEA surface after the injection of calcein-AM. Scale bar 150 µm B) 5X fluorescence microscope image to the MF MEA surface after the injection of propidium after electroporation protocol. Scale bar 150 µm C) 5X fluorescence microscope image to the MF MEA surface after the staining with DAPI.*

The picture above (Fig 15 A) shows the fluorescence image acquired using an 5x air objective optical microscope. The cell culture demonstrate to be under physiological conditions.

To allow the calcein to diffuse into the cells, the fluorescence images were acquired 1 hour after the injection in the microfluidic channels below the MF-MEA. The calcein solution was replaced with PBS, before imaging, in order to wash both the tubing and the microchannels.

As the images show, calcein is delivered only to the cells that are tightly adhered on top of the nanostructured electrodes. Strengthened also by previous work [91] [5], we could state at this point that the intracellular delivery of molecules by diffusion through hollow nanotubes does not affect cell viability.

After the rinsing of the microfluidic channels, we injected into the tubing a saline solution with propidium iodide (PrhD-1) at a concentration of 1 mM in PBS 1×. This molecule is non-permeant to the cellular membrane and when binds nucleic acids emits a red fluorescence at $\lambda = 635$ nm. In order to open the transient pores in the cellular membrane and to deliver the propidium iodide inside the cells, electroporation pulses were applied through the 3D nanoelectrodes of the MF-MEA [82]. In particular square pulses from 0 to 2.5 V with 50 ms period and 100 μs pulse duration were applied between the nanostructures and the reference electrode in the bath.

As it can be seen in fig 15B, PrhD-1 is delivered only to the cardiomyocytes lying on the electrodes that are subjected to electroporation, providing a clear evidence of the intracellular delivery achieved by the platform. This result is in line with previous results [86] and confirms the delivery efficacy of the hollow nanostructures. Furthermore, a high selectivity and spatial control is clearly reached in this experiment, because no staining is observed in cells not laying on the hollow nanostructures. The effect is due to the tight cell adhesion on the hollow nanopillars, which avoid the diffusion of the dye molecule into the rest of the cell culture.

The possibility of simultaneous delivering of both non-permeant and permeant molecules with electrical recording of these cells was also investigated. For this purpose, six devices were used. The electrical recording was performed in all of them while the simultaneous delivery was verified in 2 MF-MEAs. The Figure shows bright-field and fluorescence images of four MEAs electrodes during the injection of both calcein-AM and PrhD-1. The cells were cultured on the device as it has been explained above, when the culture reached confluence we could record its spontaneous electrical activity. The recording experiments were performed outside the incubator keeping the temperature at 37 °C by means of a Peltier cell. In order to maintain sterility without affecting the $CO_2$ exchange, a thin PDMS cap was attached to the glass ring forming a sealed culture chamber. The MF-MEA biosensor was at this point mounted on a custom-made amplifier (MEA acquisition system) acquiring data from 24 electrodes at a 10 kHz sampling rate. In fig 16A the bright field image shows the cell on the surface during the recording, calcein could diffuse through the nanoelectrode directly into the cell as seen in the previous experiment. The electrical recording does not affect the delivery of the dye via the nanostructures enabling the simultaneous injection and recording (fig. 16E). After the application of electroporation, also popidium iodide could diffuse into the cells. The protocol has been performed just on the two circled electrodes in such a way to demonstrate the localized intracellular delivery just trough the two electrodes on which electroporation was applied (fig 16C). The other two electrodes, since no electroporation was applied, the propidium could not overpass the cellular membrane and then be expressed into the cell. In fig16D it is possible to notice, therefore, the selective delivery of calcein-AM on the cells lying on the four electrodes, while propidium is delivered only to the electroporated cells lying on the circled electrodes. The cell on the other electrodes does not show the expression of PrhD-1, suggesting the integrity of their cellular membrane and thus their good health. After the electroporation, also the electrical signal changed in shape and amplitude (see fig. 16F ). In fact, as it was previously explained, after the breaking of the membrane, the typical biphasic field potential is replaced by the positive action potential shape typical of intracellular recordings (see equivalent circuit model).

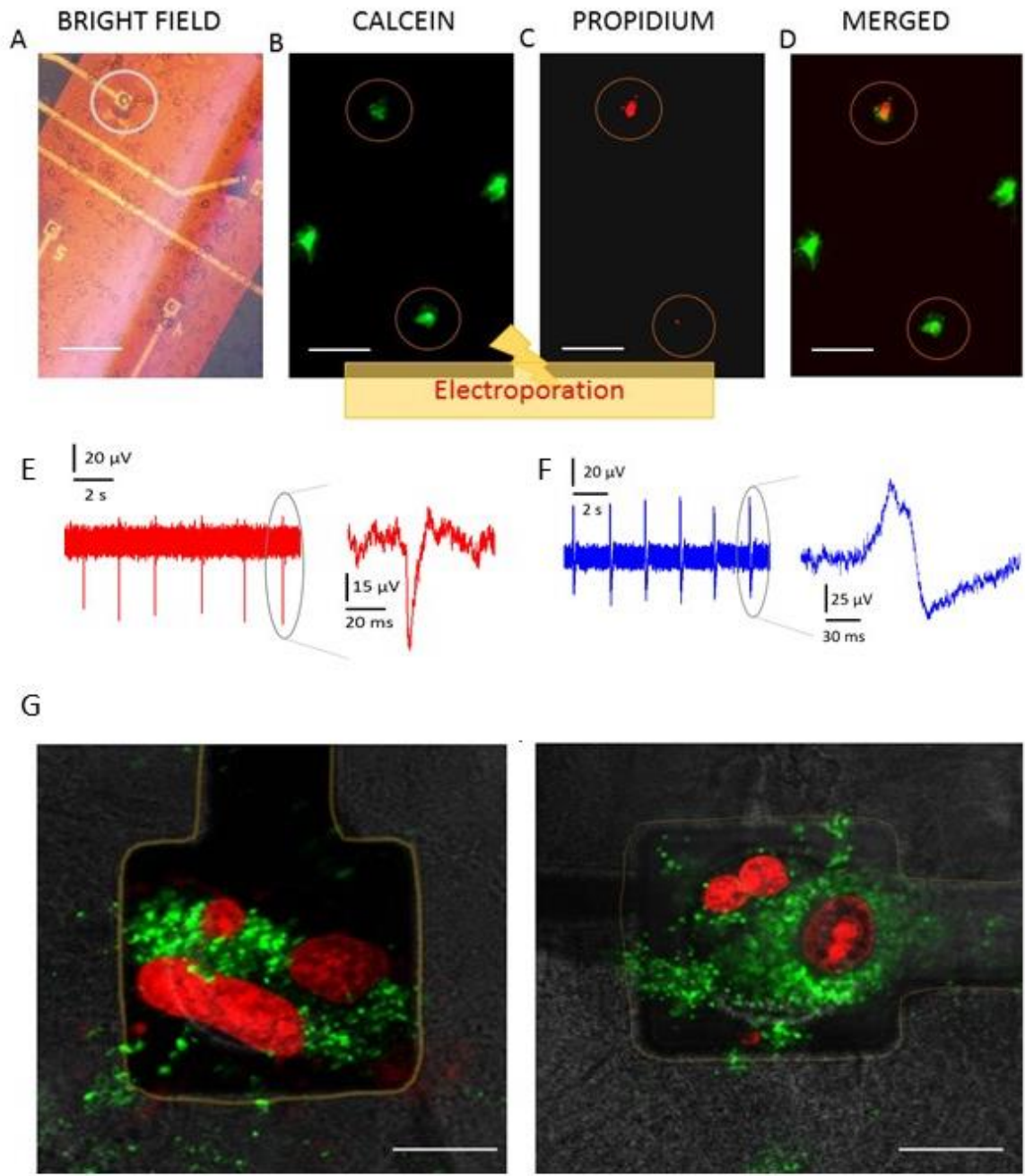

**Figure 16** *A) Bright field image of four electrodes on the microfluidic channel with HL1 cultured on top. Scale bar 200 µm B) Same 4 electrodes observed in fluorescence after the calcein-AM diffusion trough the microfluidic channels C) Same 4 electrodes after the electroporation protocol. Propidium could penetrate the cellular membrane resulting in the red colour in the image. D)Merged picture of calcein and propidium E) electrical recording on this electrodes before electroporation during injection of calceine F) electrical recording on the same electrode after electroporation during propidium diffusion G)60X fluorescence confocal images of the electrodes scale bar 10 µm.*

The cells were fixed after delivery so as to acquire high resolution images of the fluorescence emission using a confocal microscope. A magnified detail of the calcein-AM and propidium delivery on two electrodes acquired using a 60x water immersion objective is shown in fig. 16G.

# 4.5 Cardiomyocyte extracellular and intracellular spontaneous activity recording in hiPSC

Extracellular and intracellular recordings were performed on cardiomyocytes derived from human induced pluripotent stem cells (hiPSC) using the hybrid nanostructures both as sensing elements and for electroporation. The hiPSC derived cardiomyocytes represent an encouraging future standard for drug screening experiments on cardiac cells and are being investigated widely by the scientific community [92]. These cells moreover, thanks to the newly developed genome-editing techniques that allow to manipulate the genome for creating patient-specific hiPSC, represent an appealing cell source to develop disease-modeling assays, drug testing assays and cell-based replacement therapies especially in case of patient-specific care [93].

HiPSCs (Axiogenesis) have been thawed and pre-cultured in a cell culture flask and growth in Cor.4U complete medium (Axiogenesis) before seeding on samples. This procedure allows the removal of dead cells prior to seeding and will result in better assay performance. The MF-MEAs have been sterilized with 20 minutes-UV exposure in laminar flow hood and incubated the complete medium in order to saturate the PDMS passivation of the electrical connections. After the incubation, samples were treated in order to promote the tight adhesion of cells. The substrates were coated with Geltrex ready-to-use solution (Thermo Fischer Scientific) and then incubated for 30 minutes at 37°C in a humidified environment. The solution was, at this point, totally removed and cells were rapidly plated without the coating was allowed to dry. HiPSCs-derived cardiomyocytes have been plated at the density of 70000 cell/ cm$^2$ and grown with Cor.4U complete medium.

The MF-MEA device was connected with the acquisition system and the 24 electrodes could record the signals from the cells above them with the same modalities explained in the previous paragraph. Electrical recordings on the MF-MEAs were acquired using 4 different devices. The recordings were performed after roughly 4–5 DIV to obtain a spontaneous synchronized firing, allowing the culture to reach confluence. In the fig. 17A it is exemplified the spontaneous extracellular firing of hiPSC cardiomyocytes on one of the 24 electrodes of the device.

A detail of hiPSC cardiomyocyte spontaneous electrophysiological activity from one electrode after 4 DIV is shown in fig 17C. The extra-cellular potentials present high signal-to-noise ratio (SNR), allowing for the full characterization of the cardiomyocyte culture in terms of beating frequency and propagation patterns. Even for this case, the presence of the microfluidic channels underneath the device does not influence the electrical recording performance of the MF-MEAs. The extracellular spikes reveal a significantly long repolarization phase at the tail of the signal.

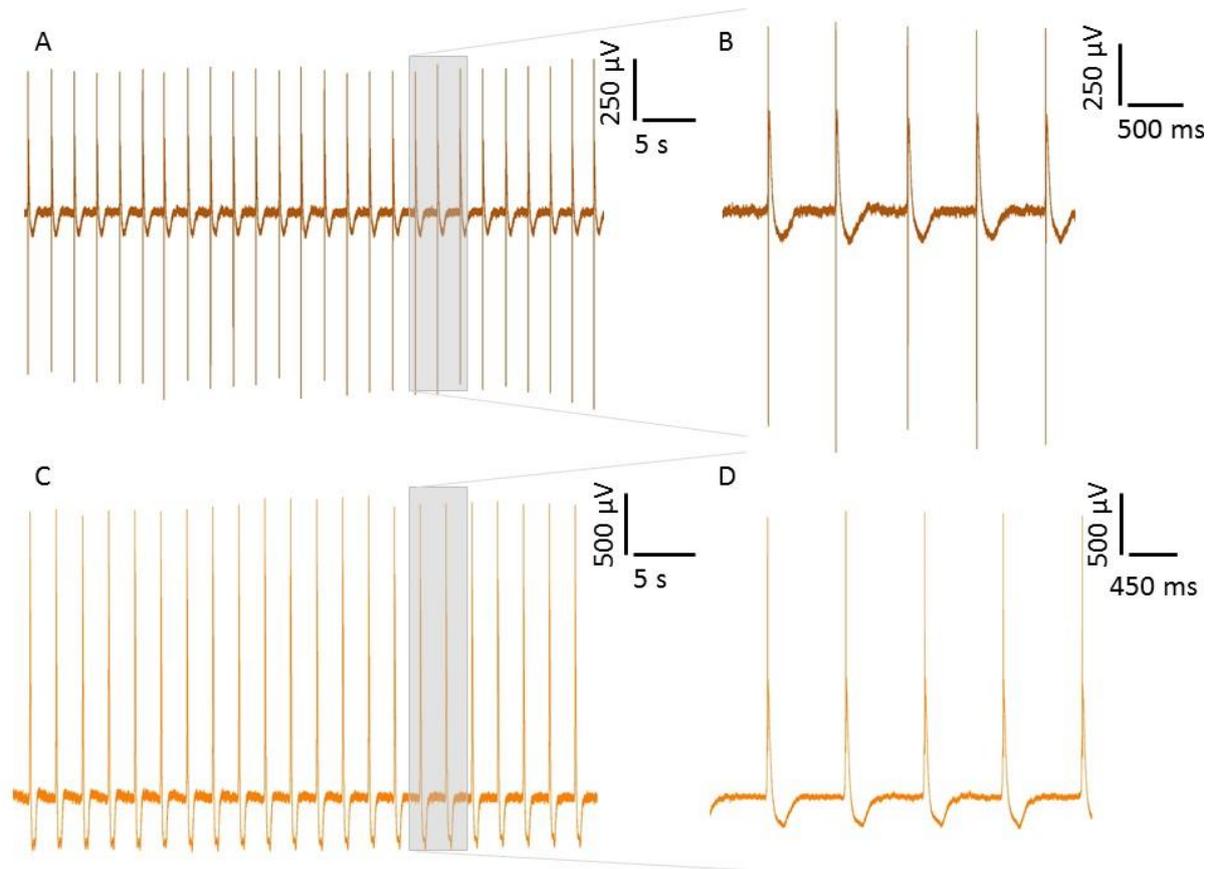

**Figure 17** *A) a short recording of extracellular signal of hiPSC B) A detail of the previous track C) short recording of intracellular action potential and D) a magnification of the signal.*

This is typical for the used human derived cardiomyocytes, which are predominantly of ventricular-type and thus have a long repolarization phase. The long repolarization phase is observed also in the extracellular mode because of the low frequency value (1 Hz) used for the high- pass filter on the acquisition board (necessary to acquire intracellular signals after electroporation).

To achieve intracellular recordings, the soft electroporation protocol was used [82]. This allows the 3D nanoelectrodes to porate the cellular membrane, applying a voltage pulse train (square pulses from 0 to 2.5 V, 50 ms period, 100 µs pulse duration) through single nano-structured electrodes, without disturbing the rest of the cell culture. The intracellular recordings exhibits (see fig. 17C) a clear change in both the spike duration and waveform, together with the typical polarity inversion. The major changes is the duration of the spike that increases from a few milliseconds to more than 50 ms after the switch to the intracellular-like configuration (fig. 18) allowing for the proper characterization of drug effect on these kind of cells.

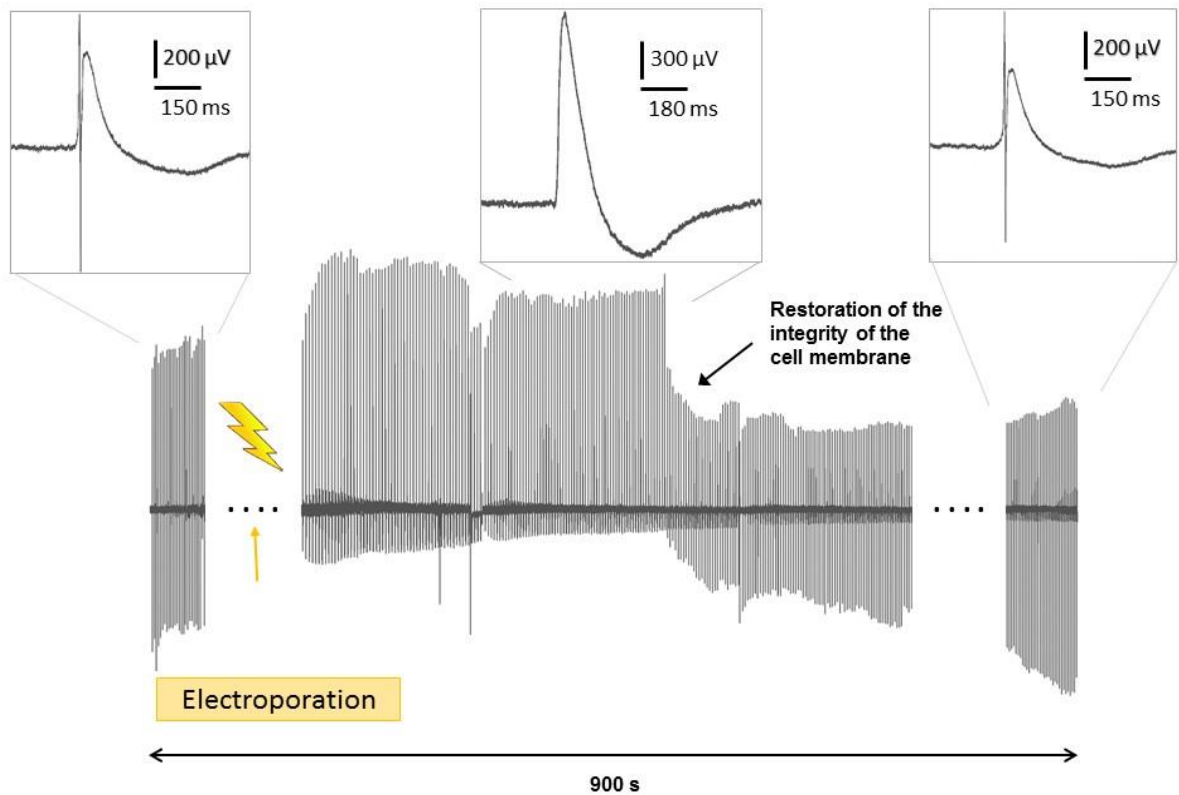

**Figure 18** *Long track recording of the cells activity before, during and after electroporation*

In fact, the effects of many treatments on the heart are mainly determined by the variations in depolarizing and repolarizing currents, which can be studied by the spike duration at different amplitudes. The intracellular coupling can last as much as 900s, with the post-electroporation signals maintaining the intracellular shape over several minutes with a soft decreasing in amplitude. This behaviour is due to the resealing of the transient membrane pores. After the membrane recomposes its natural state the spontaneous activity recovers the extracellular features. This process takes about 10–15 minutes as already observed in the literature for similar cells after electroporation.

# 4.6 Recording of neural spontaneous and stimulated activity induced by caffeine delivery

Electrical recordings of rat hippocampal and cortical neurons were performed in order to examine the response of the device with neural networks. Primary neuronal cultures from rodents are widely used to investigate physiological properties of neurons, such as development, aging and death, and represent a powerful tool to study the potential neurotoxicity of molecules [94]. These rodent neuronal networks have been used as a well establish model for studying human brain properties and diseases over the years [95].

For examining the spontaneous activity of the network, the experiments were performed outside the incubator following the same procedure and using the same setup of the previous experiments with cardiomyocytes. The neurons were cultured at a density of $10^3$ neurons/mm$^2$ on top of the MF-MEA biosensor.

The spiking activity could be recorded from the device after keeping the neurons 3 weeks in the incubator. After elongating processes and establishing synaptic connections, the cells presented a functional neuronal networks. The experiments were carried on 5 different devices at 18, 21 and 22 DIV for cortical neurons and 19 and 20 DIV for hippocampal neurons. Each recording lasted a minimum of 3 minutes and the activity was recorded several times for each experiment.

Typical waveforms of the local field potential are shown in figure 19, where negative and positive phases can be seen. The typical behaviour of the mature neuronal culture in physiological medium consists of synchronized firing and bursting activity. This last consists in an intermittent collective behaviour characterized by quiet periods sprinkled with intense spiking activity.

As depicted in the graph below, the extracellular spikes present a good signal to noise ratio (SNR) allowing for the characterisation of the bursting and firing activity of the cells on the electrodes.

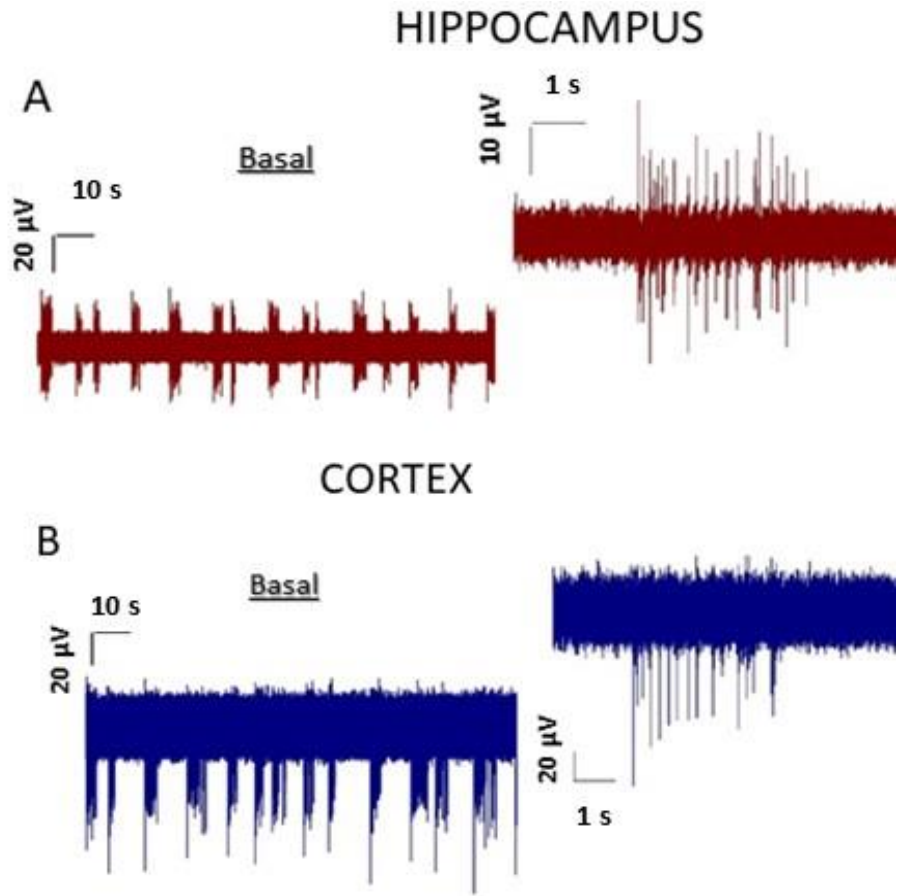

**Figure 19** *Basal spiking activity for hippocampal A) and cortical B) neurons with magnification of the burst event for both cultures*

The spike detection was performed off-line by using a custom software package named SpyCode [96] developed in MATLAB (The Mathworks, Natick, MA, USA). The acquired raw signals were peak detected by means of the threshold-based precise timing spike detection (PTSD) algorithm [97]. We also performed burst detection according to the method described by Pasquale et al.[98]. To quantify the network dynamics properties, from the obtained spike trains, the following metrics have been computed for both cortical (21 DIV) and hippocampal (20 DIV) neuronal networks: mean firing rate (MFR) and mean bursting rate (MBR).
MFR and MBR were calculated in physiological conditions. In the following table, I reported the mean MFR values representative for the recorded activity in cortical neurons at 21 DIV and the hippocampal ones at 20 DIV, both of them calculated per each active electrode. While the MBR values were the median values from all electrodes that exhibited bursting behaviour.

|  | MFR (spike/s) | MBR (burst/min) |
|---|---|---|
| Cortex 21 DIV | $0.5 \pm 0.1$ | $3.1 \pm 0.9$ |
| Hippo 20 DIV | $2.2 \pm 0.6$ | $16 \pm 4$ |

In fig. 19 it is shown this activity in both cultures for an example electrode of the MF-MEA situated on silicon nitride membranes. The blue track represents the cortical culture at 21 DIV while the red one depicts the hippocampal activity at 20 DIV.

The reported values and the recorded signals are in agreement with the literature for mature healthy rat cortical networks[99][100]. The existence of microfluidic channels underneath the device, therefore, does not affect the electrical measurement and the network behaviour, reinforcing the observations in the previous experiments with cardiomyocytes. The nanostructures and the microchannel does not influence the cells growth making the suitable to receive the cells up to the maturation state. Both firing and bursting behaviour can be recorded, allowing for a proper characterization of the network activity with the device.

### *4.6.1 Caffeine delivery*

Caffeine has multiple effects on the central nervous system, resulting also in an increasing of neuronal activity. Mobilization of intracellular calcium, inhibition of phosphodiesterases and working as antagonist of adenosine receptors, these events all together act as stimulant for the cells under examination [101].

In this experiment, the cells were chemically stimulated by localized caffeine injection through the nanochannels. In particular, caffeine solution was prepared at a concentration of 16 mM in PBS buffer, was afterwards, incubated at 37°C and then inserted into microfluidic channel with the pumping system. After filling the tubing and the compartments underneath the biosensor, the caffeine is, at this point, able to diffuse along the nanoantennas up to the level where the cells are cultured. Therefore, the consequential increasing of the activity will be enabled just at specific location of the network thanks to the platform design while the rest of the cell ensamble will not be affected. This will emulate a co-pathological state of the culture, comparing the activity of the affected electrodes with the response at the closest electrodes without access to the delivery system.

To assess the selective delivery of these molecules on the cells, a total number of 5 cultures were used. The phenomena were explored by observing the changings in firing/bursting rate of the affected cells and comparing it with the case of electrodes without hollow structures.
In the fig below is presented a typical recording after the caffeine delivery on hippocampus (fig. 20 A) and cortex ( fig. 20 B) at 20 and 21 DIV respectively. The increasing in firing activity is clearly seen in comparison with the graph (fig. 19) where no molecules where introduced.

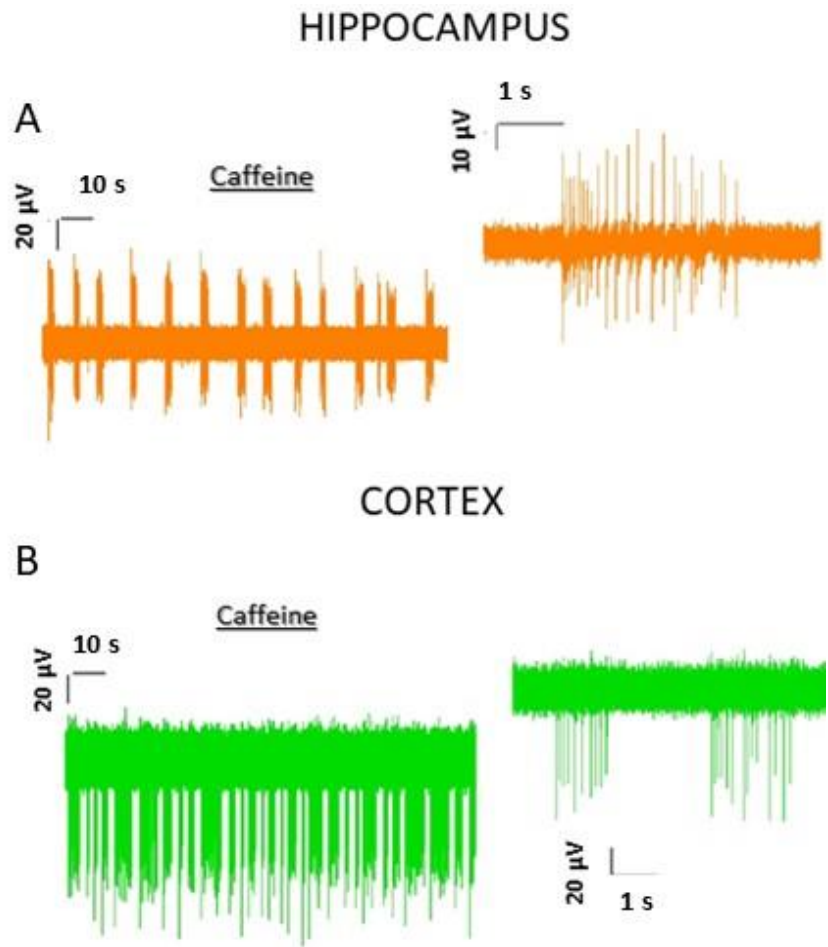

**Figure 20** *Stimulated activity after caffeine injection for hippocampal A) and cortical neurons B)*

The histograms (fig. 21) shows the difference in firing rate between the electrodes with hollow nanostructures on microfluidic channels and without on the bulk silicon nitride. The rate was calculated employing the integrated multichannel system software with the algorithm for spiking analysis. The reported values represent the percentage difference from the basal case on both nanostructured and non-nanostructured electrodes.

As revealed in the histograms, hippocampal and cortical neurons show an increasing of activity just on the cells in correspondence of the nanostructures. Therefore, caffeine was well delivered through the microfluidic channel without affecting the cells on planar electrodes with no direct access to the microfluidic compartments.

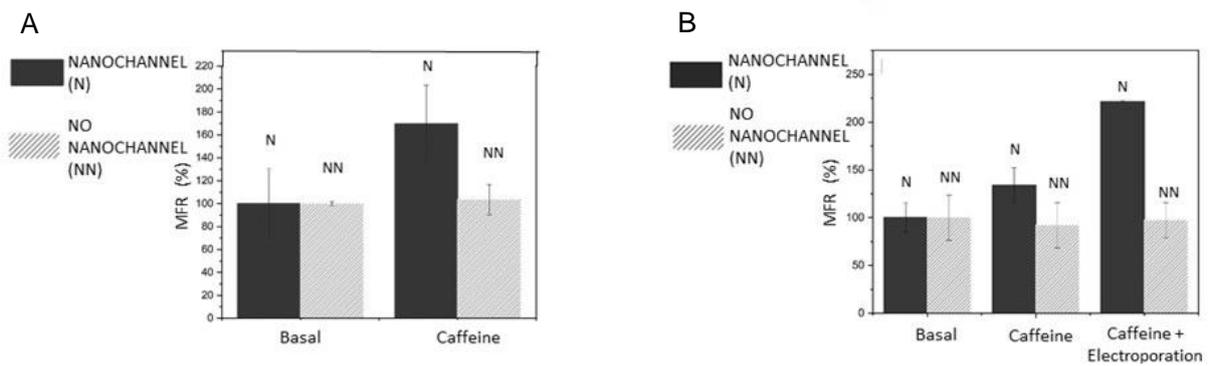

**Figure 21** *Histrograms representing the firing rate of hippocampal A) and cortical B) neurons under physiological and stimulated (caffeine) condition. The values represent the percentage discrepancy from the basal case. The full color represent the electrodes with nanostructures on top, while the dashed one represent the pads without.*

As a further trial to exploit the capabilities of the device, an electroporation protocol was applied on the cortical neuron during the caffeine injection [82]. With this protocol, we were able to just partially permeabilized the cellular membrane, leading to a further increasing of the firing rate as observed during the experiments. The results are shown in the histogram (fig. 21B).

Given the obtained results, the multifunctional platform demonstrates to be suitable for primary neuronal cultures recording both in basal and modified conditions. The device enables the delivery of stimulant molecules, such as caffeine, in the specific area of the electrode without diffusion on the cellular medium enabling for high-localized drug delivery studies.

The conservation of the electrical activity after electroporation and the caffeine delivery ensures that the cells are not damaged by these procedures.

# 4.7 Conclusions

A novel multifunctional platform was presented with integrated microfluidics and recording capabilities. The approach combines cutting-edges nanotechnologies for intracellular delivery and intracellular recordings. The geometrical hollow configuration of the nanostructures supports a precise delivery on the few cells located on the electrode area without disturbing therefore the collective behaviour. For the first time we could see this achievement implemented on a MEA in-vitro device.

The device have shown to work with three different cellular types. The robust and standard model for electrophysiological studies represented by HL1 cell line, the *in vitro* cell model that could reproduce as closest as possible the physiological human environment (hiPSC) and as last the primary neurons that represent a high reliable model for neurologic studies. The high quality intra and extracellular recording of these cell lines demonstrated that the device could be suitable for both cardiological and neurological studies.

We successfully demonstrated intracellular delivery over HL1 cell line during extra and intracellular recording. The delivery was limited to the area upon each electrode ensuring the high localization of the molecule. The intracellular delivery could be enabled just at one electrode site avoiding cross talk between the 24 electrodes and allowing the injection of the macromolecules just on that specific site.

A stimulant molecule was specifically injected to the primary neurons in order to test the effect on the network activity[102]. The delivery precisely reach just the neurons in correspondence of the electrodes on the microfluidic channels and the molecules does not diffuse on the rest of the culture, reinforcing the already demonstrate selectivity provided by the device. The high recording capabilities allowed to discriminate the change in activity due to the caffeine injection.

The presentation of the Microfluidic MEA with the fabrication process, the recording of HL-1 and hiPSC derived cardiomyocyte together with the demonstration of selective intracellular delivery has been published in "*Selective intracellular delivery and intracellular recordings combined in MEA biosensors*" [103]. While, the detailed results on neuronal culture could be visualized in a further publication that is under submission (see Publications).

## 4.8 Perspectives

This technology appears very promising as a tool for pathology studies, drug development or basic biology understanding. However, the requirement for 3D structures nanofabrication limits the applied implementation of this platform as a large scale and easily accessible research tool. The main reasons are attributable to the complex and costly procedures that are not well suited to mass production. The 3D nanostructures fabrication protocol requires techniques that in term of cost and time consumption are too high for commercial scale.

The use of plasmonic metamaterial could overcome this limitation, as will be explained in details in the next chapter (Porous plasmonic material). The extra and intracellular recording capabilities of porous plasmonic electrodes have been demonstrated in previous works [10], together with possibility of membrane poration. For our purposes, the porous electrodes will be fabricated on the silicon nitride membranes allowing for the selective intracellular and extracellular drug delivery. The combination of these characteristics on the same device will allow us to reach the discussed result. For sake of clarity, the concept and biosensor design is depicted in fig 22.

This process, thus, could be suitable for the large scale production of MF-MEAs. Moreover, beside reducing cost in fabrication, it will allow to a higher standardisation of the electrodes manufacturing ensuring higher quality devices and reproducibility. The number of electrodes could be easily increased improving the resolution of the device.

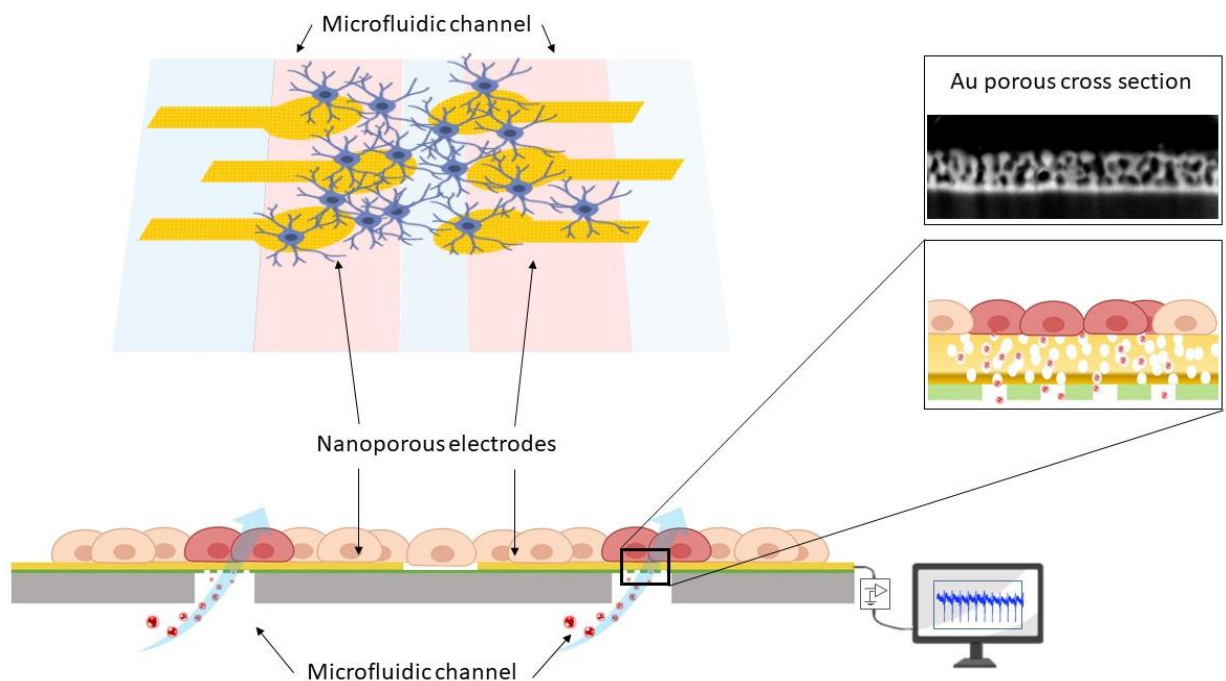

**Figure 22** *Schematic of porous microfluidic MEA with SEM image of a cross section of fabricated nanoporous gold on silicon nitride membrane.*

# 5 Plasmonic porous materials

## 5.1 Possible perspectives

*Electrical stimulation of excitable cells*

Electrical stimulation of excitable cells is the basis for many implantable devices in cardiac treatment (such as the pace maker) and in neurological studies for treating debilitating neurological syndromes[104][105][106]. Traditional stimulation electrodes (as multielectrode arrays), as broadly explained in section ([Multielectrode arrays](#)), allow to interface and stimulate cardiac cells and nervous systems. However, the electrical stimulation presents multiple challenges, such as selectivity, spatial resolution and mechanical stability, implant-induced injuries and the consequential inflammatory response.

Alternative methodologies, such as optical approaches, are widely investigated because of the intrinsic low invasiveness and single cell targeting, while also permitting to avoid the electrical artefacts that complicate recordings of electrically stimulated neuronal activity.

Light has been known to influence the behaviour of electrogenic cells since the work of d'Aarsonval in 1891 [107]. R. L. Fork [108] showed that abdominal ganglion neurons in *Aplysia californica* respond to 488 nm laser light, through a reversible mechanism, despite the cells are not photosensitive. Building on this work, other groups demonstrated how laser irradiation promoted membrane depolarisation and action potentials in spontaneously active neurons [109][110]. Following this trend, optogenetics is considered one of the major achievement for addressing optical stimulation. However, it needs for genetic modification of the cells in order to introduce photosensitive receptors in the neurons, being thus a highly invasive technique [13]. Over the past decade, a number of new techniques have been developed in order to use light to trigger electrical responsive cells. Dissimilar to optogenetics, the *non-genetic* photostimulation does not change the biological framework but uses transient thermal or electrochemical output from synthetic materials attached to the target cells.

In the field of *non-genetic* photostimulation, the photo-thermally modulating materials meet the requirements of being minimally invasive, integrated with the target biological system. However, the chronic effect due to the heat of these devices are still unidentified [111]. Aside from these platforms, quantum dots, gold nanoparticles and semiconducting polymers have been widely utilized for optical neuromodulation in retinal photostimulation, mouse hippocampal brain slices and for restoring light sensitivity in blind rats [112][113][114][115]. Silicon nanomaterials, on the other hand, have shown optimal responses in the field [116]. The group of Bozhi Tian demonstrated how silicon nanowires, thanks to the highly tunable chemical and electrical properties could be used as promising tool for neural modulation[15][117].

In these processes, the physiochemical events occurring at the interface between the stimulated material and the electrolyte (cell medium) are due to the charge transfer between the material and ionic solutions [118]. Charge transfer can occur through three mechanisms: (I) non-faradaic charging/discharging of the electrochemical double layer, (II) reversible faradaic reactions, and (III) non-reversible faradaic reactions. The latter should be avoided because it is associated with electrode degradation and harmful products. The first two mechanisms are due to the always-present charged layer between the electrode and the solution (see section on the double layer). Studying these two phenomena could give an indication of the capabilities of the material to redistribute or to inject charges into the cell or tissue. The faradaic process, characterized by electron transfer between the electrode and the electrolyte, is particularly constructive in biological systems. Charge injection has a fundamental role in many biological processes where the electrons participate as a trigger in stimulation of electrogenic cell.

Bozhi Tian and co-workers stressed how Faradaic current has implication in eliciting neurons [14]. Since, the cellular physiology could be altered by approximately a picoampere level ionic current, it was shown that the produced photocurrent was able to efficiently elicit cellular assemblies or rat brain slices.

According with the results shown in the aforementioned works, plasmonic meta-materials could be exploited as a cutting-edge technology in cardiac activity stimulation.

In this context, P. Zilio et al. presented a theoretical and experimental study on a plasmonic nanoelectrodes architecture that clarifies the injection of electrons from the structures into aqueous environment [16]. The results suggest that the 3D structures, connected to flat electrodes, could provide an infinite reservoir of electrons allowing the long-term generation of hot electrons. More in details, the laser-excited nanoelectrodes in water offer a suitable way to produce 'hot' electrons for photocatalysis and electrochemistry. When the gold nanoantennas is illuminated with ultra-short laser pulses, plasmonic hot spots are formed on the nanoantenna tips. This phenomenon enhanced multiphoton absorption and the consequential injection of electrons into the water. Accelerated by the strong plasmonic fields, these injected electrons attained kinetic energies of tens of electron volts, which are sufficient to ionize water and generate electron avalanches.

Considering the charge injection capabilities of the 3D counterparts, the use of plasmonic metamaterial could be a farsighted option for developing a low cost, non-invasive, optical stimulation device for *in vivo* and *in vitro* applications. Moreover, the method could be easily implemented on standard CMOS MEA device. In fact, tuning appropriately the optical parameter, the technique presented in [10] could be used for cardiac stimulation at single cell level on high-density and high-resolution sensors.

In the next paragraph, experiments performed on nanoporous gold will be discussed for demonstrating this hypothesis.

## 5.2 Applications

During the last decade, there was a great interest in exploring the potential of nanoporous materials for sensing. In particular, the metallic ones, thanks to their intriguing plasmonic properties, show high potential in different fields from electrochemistry [119] to nanofluidics [120] to optical biosensing in applications [121] such as surface plasmon resonance (SPR), sensing and surface-enhanced Raman scattering (SERS) [8] and MEA technology[89][122].

The SPR spectroscopy is a well-known detection method useful to monitor the surface binding events in real time, which gives rise to the changes in the dielectric environment at the surface. The traditional SPR method makes use of the properties of surface plasmons in the form of propagating surface plasmon polaritons (SPPs). This method is difficult to be integrated into portable, low-cost devices and high-throughput systems[123]. The SERS system, on the other hand, was widely applied for analytical detection with high sensitivity and low detection limit. This technology offers the opportunity for developing platforms for high enhancement in Raman scattering signals of the analyte molecules adsorbed on the SERS substrates. A number of attempts have been made to fabricate high-performance and reliable SERS substrates, for examples, engraving periodical nanosphere arrays by lithography or self-organizations of nanoparticles in the solutions [124].

Materials such as nanoporous gold (NPG) offer a valid alternative providing low-cost fabrication methods and simple excitation schemes of strong plasmonic resonances, in all the above-mentioned applications. Among the several newly investigated plasmonic materials, NPG presents significant advantages mainly due to excellent thermal stability, chemical inactivity and the large surface/volume ratio[125]. Nanoporous gold behaves like a metamaterial whose effective dielectric response can be tuned by changing the pore size and porosity, or in the same way by changing the fractal dimension[126]. In particular, the plasma frequency can be shifted over a wide spectral range of infrared wavelengths.

These properties, combined with a higher skin depth in the order of 100–200 nm, enable the penetration of light deep into the nanopores where the analyte can be accumulated. Such an efficient co-localization of analytes and optical energy promotes a robust light–matter coupling and a high detection sensitivity. Moreover, the topography of the material ensures higher available surface in order to use it as efficient nanoelectrode, along with the high roughness that is well known to promote a tight cell adhesion and better electrical contact[127].

Porous materials were exploited in the last years by my group thanks to their plasmonic characteristics in different fields. Mainly, in the field of electrophysiology, this material opened up the way to allow commercial multi-electrode arrays to record intracellular action potentials in large cellular networks[10].

In fact, planar porous electrodes in combination with plasmonic optoacoustic poration can mimic the behaviour of 3D plasmonic nanoantennas. More in details, in the past years De Angelis et al. largely showed that 3D plasmonic antennas excited with near-infrared (NIR) light can be used to generate mechanical nanowaves able to locally open transient nanopores in a cell membrane that is in tight contact with the antennas[86][51]. This process is called optoacoustic poration and it is different from conventional laser poration in many aspects. This transient poration, in fact, is induced by a mechanical wave (generated by the optical pulse) and the process generates only a negligible transient temperature, but it does not produce heat accumulation or other side effects. For this reason the technique can be considered minimally invasive for intracellular recording or drug delivery, and it was demonstrated to represent a valid alternative to electroporation.

Simple layer of nanoporous metal presents a comparable plasmonic behaviour under NIR light illumination. A thin layer is able to collect most of the incident light efficiently and transfer the optical energy into the nanogaps of the material in which the plasmonic field is strongly amplified. The planar porous electrode, thus, will emulate the optical and biological behaviour of three-dimensional plasmonic nano-electrodes without the drawbacks related to 3D fabrication. M. Dipalo and co-workers demonstrated how these materials could be used as a plasmonic meta-electrodes suitable for cardio-toxicological studies if implemented with fast scan laser systems. The technology is suitable to be used with commercial CMOS-MEAs, making the method robust, reliable and easy to access and use[10].

Parallel to 3D nanoantennas, also meta-electrodes provide an optimal coupling with external electromagnetic radiation in the way that the optical energy is focused exactly were the cells membrane is situated. Significantly, the coupling of the membrane with the porous surface allows for preserving, after membrane optoacoustic poration, a suitable cell sealing for a high-quality intracellular recording. The laser excitation is a highly localized process that does not affect the performance of the remaining electrodes, allowing for simultaneous poration and recordings.

# 5.3 Characterization of the photocurrent generation from plasmonic meta electrodes

The photoresponses of porous gold could be exploited by studying the phenomena occurring at the interface when the porous material is illuminated with ultra-fast laser pulses. In order to reach this goal, the following preliminary experiments were carried out. A gold porous multielectrode array was fabricated and the laser beam was aligned with a single electrode at a time. The electrode-electrolyte system is formed after PBS solution is inserted into the glass ring forming the well.

The device was connected to the patch clamp amplifier (Axopatch 200B, *molecular devices*) in order to record the changing from the equilibrium conditions when charges are injected into the electrolyte.

### *5.3.1 Fabrication of the porous electrodes*

During the last decade, nanostructured porous gold films have been prepared following different methods [9][125]. In this thesis, the preparation of NPG films followed a simple well-established procedure developed in previous works [126][128].

The fabrication protocol of the meta-electrodes includes few simple steps. Porous gold layer could be easily implemented in the standard MEA fabrication protocols. For the purpose, a $Si_3N_4$-Si-$Si_3N_4$ wafer was used as a substrate. Following the first steps explained in (Electrodes fabrication) feedlines and pads are designed through photolithography. A thin adhesion layer of 7 nm of titanium followed by another layer of 7 nm of gold are deposited on top of the substrate by means of electron-beam evaporation as adhesion layers ( Fig 23) . An alloy of silver/gold is then sputtered on the gold layer at a rate of 20 nm min$^{-1}$. The lift off process allows the definition of the final MEA design. The so-constituted MEA is now immersed in nitric acid solution (50% in deionized water) for 1 h, during which silver is etched away selectively, resulting in a highly porous gold structure. In the last step, the samples are rinsed out in deionized water and dried with nitrogen.

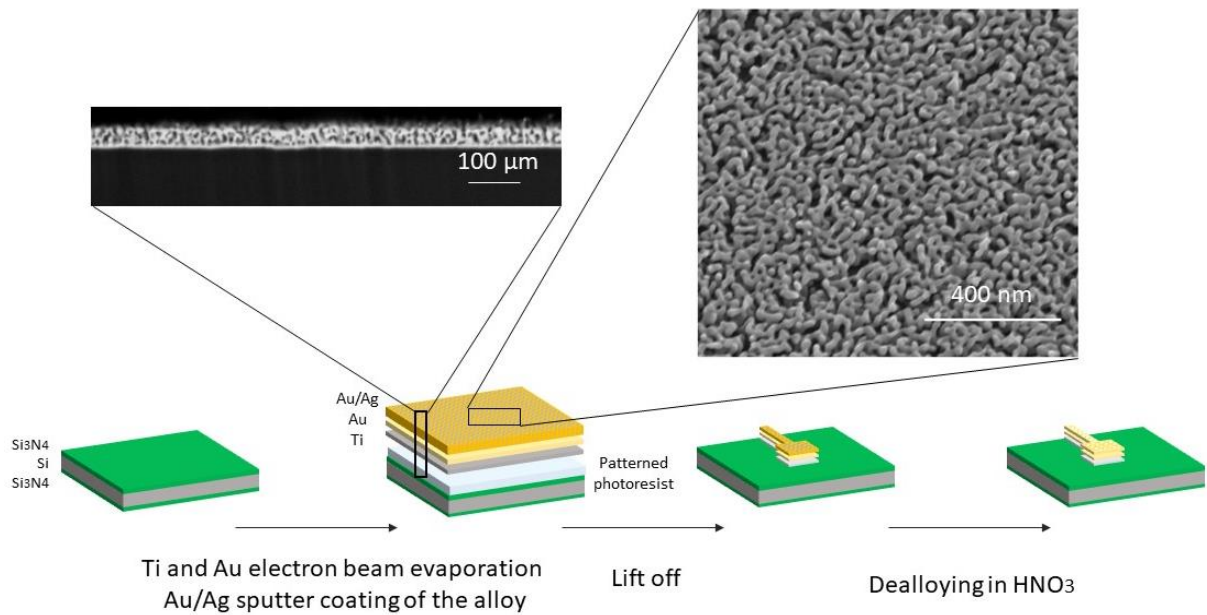

**Figure 23** *Fabrication process steps of porous MEA electrodes and SEM images of the porous gold in 45° degree tilted angle configuration (scale bar 400 nm) and cross section (scale bar 100 µm)*

After cleaning the sample, the whole surface except for the electrodes and the external pads, is passivated with SU8 in order to avoid leakage while the device is in solution. The MEA is than mounted on a PCB for interfacing with the acquisition system.

In the fig 23, a SEM image of the porous material is shown.

### 5.3.2 *Experimental setup*

In order to stimulate the photoelectrochemical reactions at the interface, the nanostructured material has to be excited with light. The 1064 nm (Nd:YAG (neodymium:yttrium–aluminium–garnet) solid-state laser (Plecter Duo (Coherent)) is used as the light source, for which the emission is in ultra-short pulses at 8 ps
with 80 MHz repetition rate. The pulsed beam is then switched ON and OFF at the desired pulse length, generating pulses train ranging from microseconds to hundreds of milliseconds.

From now on, the term pulse length refers to the ON time of the 8ps-pulsed laser, defined with an acousto-optic modulator (AOM) or a mechanical shutter controlled by a TTL signal from the analog-to-digital signal converters (ADC/DAC Axon™ Digidata® 1550B plus HumSilencer) connected to the software (Axon pCLAMP).

The laser is focused onto the porous electrodes by means of a 60× immersion objective (NA=1) using a CCD camera Nikon DS-Fi2 for visualizing the electrodes.

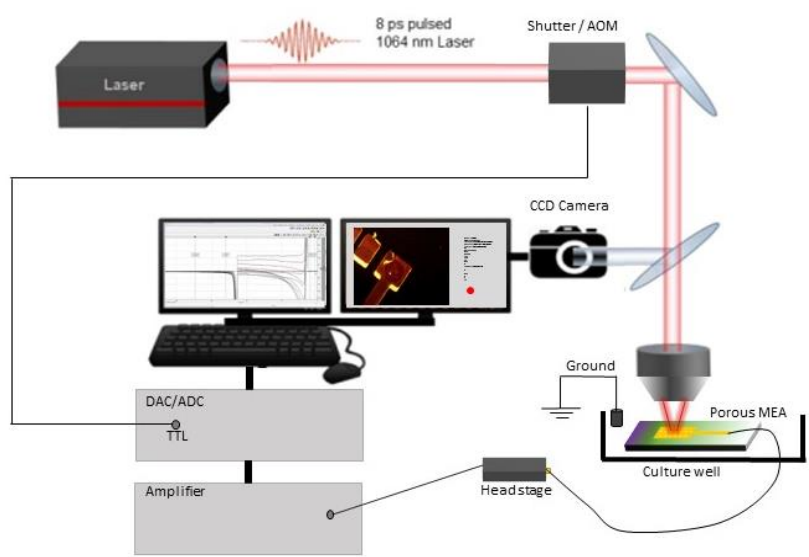

**Figure 24** *Schematic of the experimental setup. 1064 nm laser optical path that focused the beam on the porous substrate. The amplifier and the digitizer record the current at the electrode. CCD camera is connected in order to align the system.*

The porous MEA electrode is immersed in PBS solution and its external pad is electrically connected to pre amplifier (CV-7B headstage) and amplifier (Axopatch 200B for ultra low-noise single-channel recordings) in order to record the charge injection. The ADC (Axon Digidata 1550B plus HumSilence) is than connected to the amplifier in order to both send TTL signals for light pulse control and to interface the amplifier and the computer.

A rotating polarizer was used to change the laser intensity while the current at the electrode-electrolyte interface is measured. The laser intensity was measured at the beginning of each experiment by means of a power meter situated at the working distance of the objective.

A platinum wire immersed in the solution acts as counter-electrode for the current measurements; all measurements of photocurrent were made without the application of a bias between the platinum counter-electrode and the sample.

The SU-8 passivation on the sample ensures that there are no leakage currents between the platinum wire, the solution and the other gold surfaces on the sample.

## 5.3.3 Photocurrent generation of plasmonic meta electrodes

Following the above-described experimental setup, the sample was placed under the 60X water immersion microscope objective and the MEA well was filled with PBS solution. The TTL signal, set with Clampex (the commercial program from *Axon instruments*), controls the shutter in order to easily change the laser pulse duration and to decide the starting time of the pulse.

The laser spot was aligned with the porous electrode. In the picture below is depicted an image of the laser spot on the electrode.

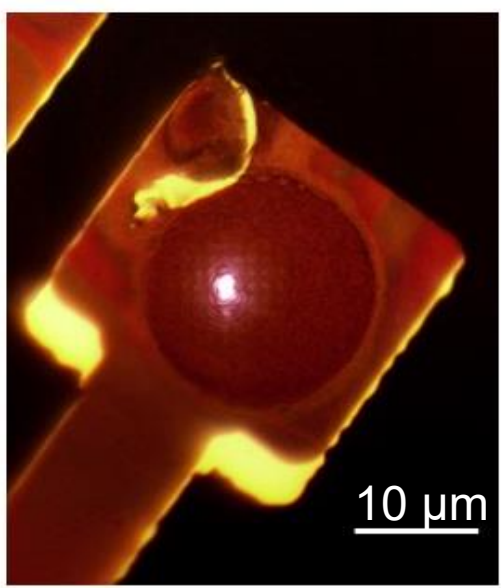

**Figure 25** *Laser spot images recorded with the Nikon camera focused with 60X immersion objective*

The measurements were performed in voltage clamp mode (set in the amplifier), holding the voltage at zero.

The measured currents were characterized by a sharp and fast initial rise, an intermediate positive shoulder and a last negative peak (that in reverse mode resembles the initial phase) (fig. 26C). As explained in the previous section, this particular shape is typical of a mixture of capacitive and faradaic responses of the system. The spikes rise and fall correspond to the starting and ending points of the laser pulse. This behaviour is typical of a capacitive charging and discharging process occurring at the interface. The positive sustained current, instead, corresponds to the faradaic component suggesting an injection of electrons into the solution resulting in electrochemical processes at the interface between electrode and electrolyte.

In order to further assess the importance of the surface morphology in obtaining this behaviour, the laser was also moved to a flat electrode, and we could observe that the laser pulse does not produce any effect on flat gold surface (fig 26 A).

The faradaic component was investigated in a first moment keeping the laser pulse duration constant, and changing the laser power. For each laser intensity set with the polarizer, the generated current was measured with the amplifier in multiple sweeps that lasted 250 ms each (example with 1 ms of laser pulse length in fig 26C). The laser spot was moved all over the available electrode surface in order to take into account the possible topological variability. Each presented current profile represents in this way the average values obtained by mediating the recordings of the multiple sweeps on the same electrode with the same laser power. We could notice an increasing of both capacitive and faradaic components as the laser power rises. This dependency of the generated photocurrent was observed at different pulse lengths. For a laser pulse duration of 1 ms (as depicted in fig. 26C) , the faradaic photocurrent were measured to be 67 pA, 197 pA and 644 pA for laser powers of 0.28 mW, 1.46 mW, and 4.86 mW respectively. The faradaic current increases following an exponential trend with respect to the impinging light power (figure 26D). A comparable current behaviour can be found for 3D plasmonic nanoantennas in [16], confirming the similar nature of the phenomenon. The maximum reported value represent the intensity at which the cavitation bubbles due to the high energy start to form. After this value, further investigations would have been misleading.

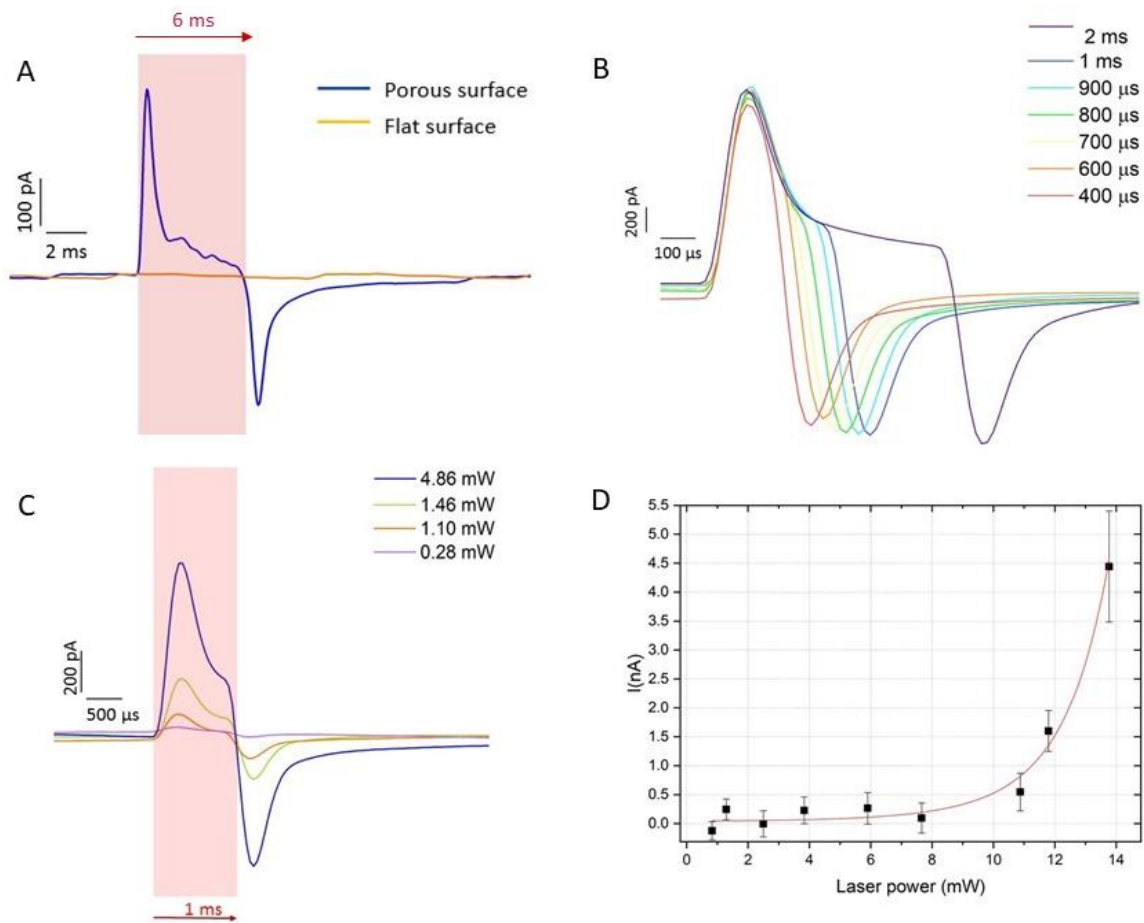

**Figure 26** A) *Comparison between photocurrent produced in the case of porous electrodes and in case of gold flat surface with 6 ms laser pulse B) Faradaic current profiles induced with different laser pulses length at the same power of 5mW C) Current profiles recorded at 4 different laser powers impinging on the porous surface with the pulse length of 1 ms D) Exponential fit of the faradaic current with respect to the laser power.*

In a further experiment, the pulse length was changed while keeping the laser intensity at 5 mW in order to study how the stimulus length affected the measured current. As represented in fig 26B, the value of the faradaic photocurrent does not increase or decrease while changing the pulse length. The observed faradaic component could be measured at pulse lengths longer than 600 µs, since the charging/discharging time constant of the double layer hided the faradaic component at shorter pulse durations. The duration of the laser pulse does not affect the amplitude of the faradaic photocurrent. The current profile suggests a mixture of faradaic and capacitive behaviour with no thermal component, as advised by previous works [14]. The recorded current values are comparable with those used in literature for stimulating electrogenic cells [129]. Thus, the porous gold seems to be a promising material for electrodes dedicated to photoelectrical stimulation. In the same way, the porous platinum electrodes, furnished by commercial CMOS-MEAs, have proven to have a strong plasmonic behaviour [10]. Therefore, in the next paragraph, the capability of plasmonic porous materials on eliciting cardiac cells will be investigated using the commercial CMOS devices.

# 5.4 Optical stimulation of action potentials in cardiomyocytes

In the presented preliminary experiments, cells from the HL-1 line were cultured on a commercial CMOS-MEA. Prior to cell plating, the device was sterilized and treated five minutes with poly-L-lysine (Sigma-Aldrich) in order to increase cell adhesion. Subsequently, HL1 cells have been seeded at the density of 84000 per cm$^2$ and grown with Claycomb culture medium. When confluence was reached (after 5 days), the electrical activity of cells has been recorded with the BioCAM acquisition system from 3Brain AG. Approximately 10 min of unperturbed extracellular activity was recorded to characterize the culture. Using the same optical setup presented in previous works [10], the laser is coupled to a modified upright microscope (Ecplise FN-1 from Nikon) able to accommodate the BioCAM acquisition system directly on the microscope stage (see image below fig. 30). A 60× water-immersion objective (with N.A.= 1.0) was inserted in the cell medium during the experiment in order to observe the cells on the CMOS-MEA electrodes and to focus the NIR laser used for stimulation.

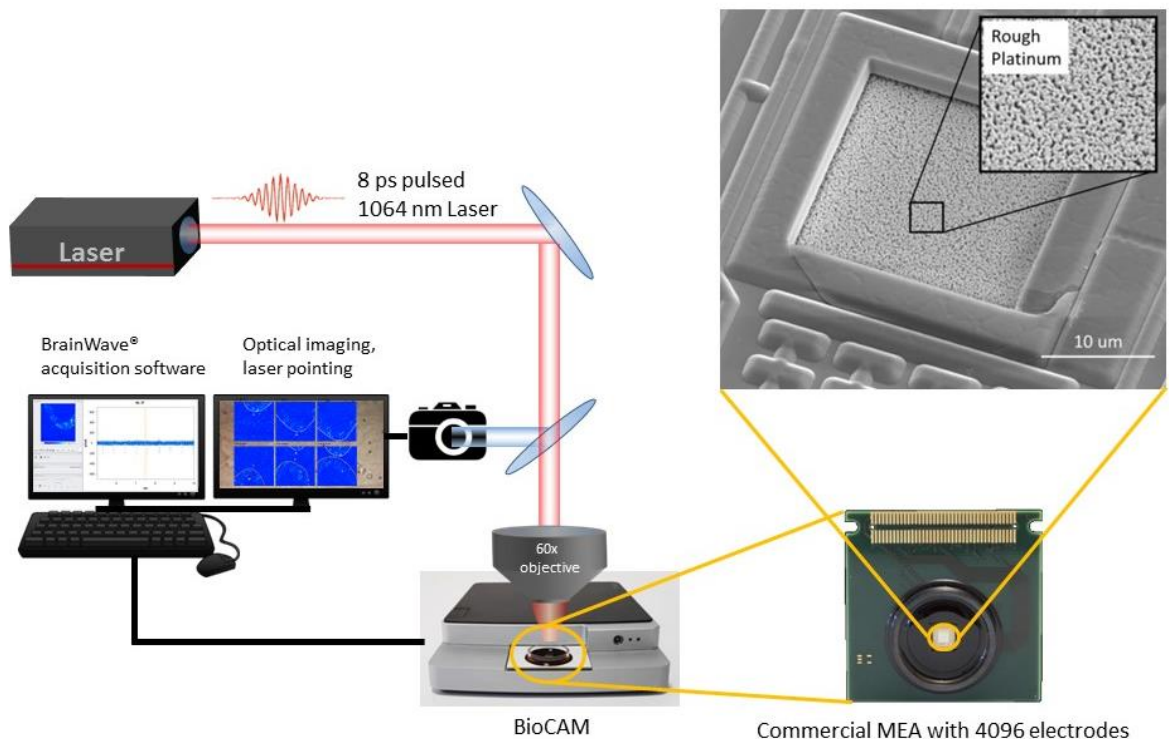

**Figure 27** *Experimental setup, commercial acquisition system and porous platinum CMOS MEA. Adapted picture from* [10]

The high resolution of the BioCAM acquisition system allows for mapping precisely the propagation waves of spontaneous activity in the cardiomyocytes monolayer. In fact, the CMOS-MEAs offer 4096 electrodes in a 2 x 2 mm area, with a spatial resolution of 42 μm.

In physiological conditions without optical stimulation, we normally observe a regular electrical activity with typical waves propagating in defined directions (fig 28A). In figure 28, instantaneous maps of the CMOS-MEA array are presented. Each pixel is representative of one single electrode, the colours range from blue up to red accordingly to the amplitude of the recorded potential variation.

The small yellow square represents the electrodes used for laser stimulation. The laser was focused on the highlighted electrodes with a tuned power of 5 mW and a pulse length of 150 ms. As in the presented previous experiment, the polarizer was used to change the laser intensity. The laser intensity was measured at the beginning the experiment by means of a power meter situated at the working distance of the objective.

In the images, it is possible to see the standard pattern and the stimulated ones from bottom and right sides of the device. In several cases, we observe the generation of a single propagation wave that originates from the stimulated cell. This is depicted in two examples in figure 28 B and C. In the fig 28B, the white highlighted wave from the top-left corner is the original propagation pattern of the culture; whereas the orange highlighted wave is a secondary propagation signal that originates from the stimulated cell at the bottom of the map (orange square) immediately after stimulation. The round region around the stimulated electrode is red due to saturation caused by the laser excitation; however, as we have seen in figure 28B, this effect does not prevent recording from these electrodes.

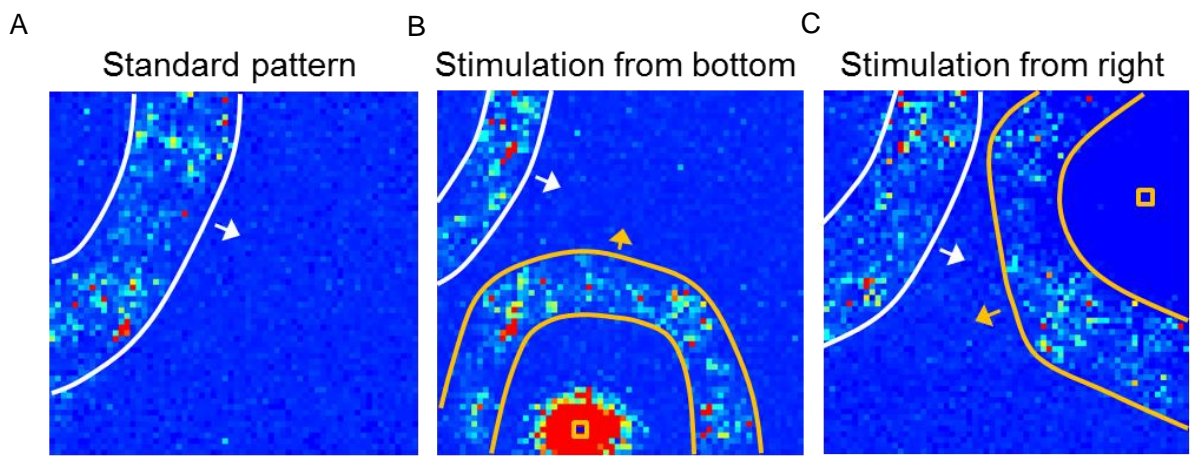

**Figure 28** *A) Standard propagation, B) and C) Change of the propagation pattern following a stimulation event triggered from a single electrode (the orange square).*

Interestingly, in few examples, we were able also to create a temporary persistent alternative propagation pattern, which is repeated more than once after the laser excitation is removed. This could be seen in fig 32 where 6 screenshots from a recording of the 4096 electrodes after the stimulation are reported. From t = 0.7s to t = 1.2s we observe the original propagation wave starting from the top and moving downward (three top panels in figure 29). At t = 1.3s we perform laser stimulation on the electrode that is highlighted with an orange square in the bottom left panel. More than 1 second after laser excitation of the electrode, from t = 2.5s to t = 3.1s, we notice that there are still two different propagation patterns running simultaneously, the original wave from the top and an alternative wave starting from the stimulated cell and moving upward (three bottom panels in figure 29). The newly generated pattern lasts for one or two cycles, then the cell monolayer restores the original downward propagation pattern from the top.

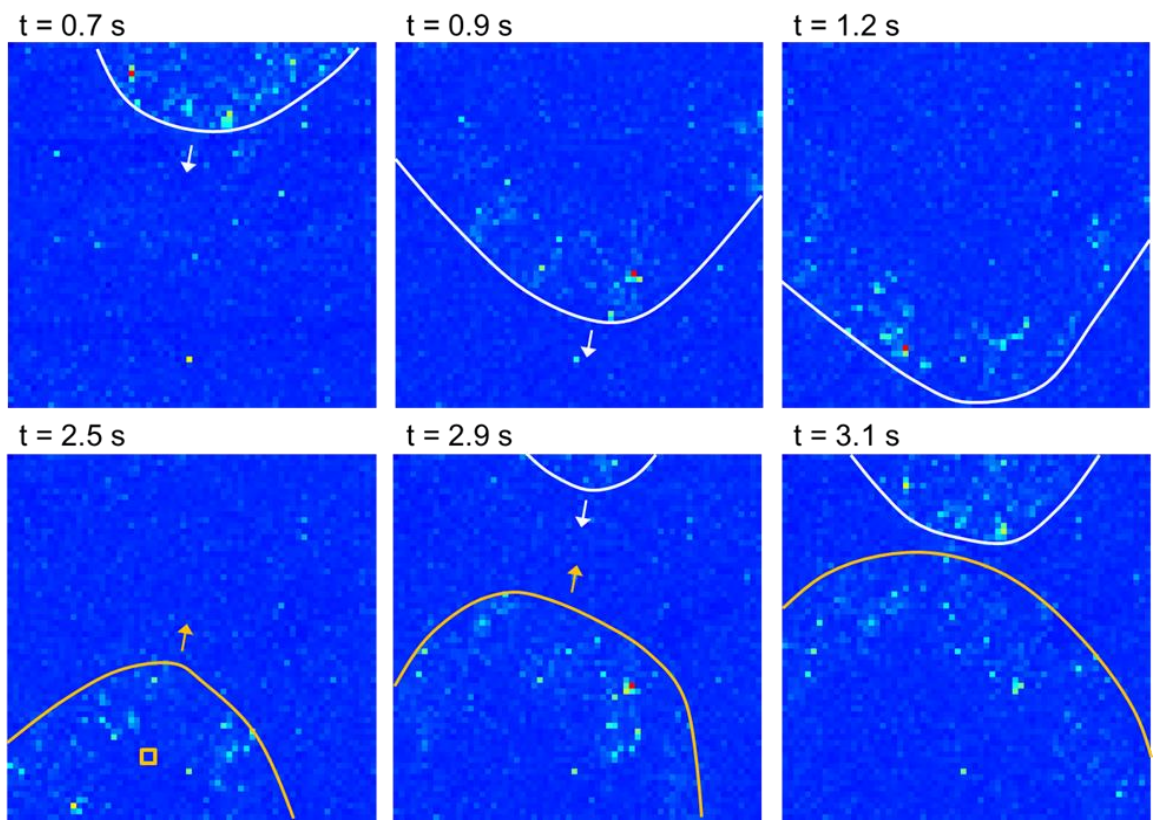

**Figure 29** *Time frame of stimulation of a persistent alternative propagation pattern from the stimulated electrode (the orange square in the bottom left panel indicates the electrode that has been excited with laser pulses)*

In summary, in this configuration, we found that photoelectrical current produced by the interaction between porous electrodes and impinging laser could elicit cardiomyocytes at a laser power ranging from 2-3 mW up to 5-7 mW, with laser pulse duration of approximately 150 ms. This is just a preliminary trial, but suggests that the photoelectric stimulation of cardiac tissue is possible using plasmonic meta electrodes. Further analysis will be carried out in order to define optimal parameters and energy thresholds in order to elicit cardiac cells.

Moreover, we found these results using commercial CMOS-MEAs without modifications of the devices. The proposed approach thus offers the possibility to expand the capabilities of the present commercial devices without compromising their performance.

## 5.5 Conclusions

In this chapter, I presented preliminary experimental data that describe electron injection in electrolytic solution by photoexcitation of porous plasmonic materials. The injection was experimentally monitored by measuring the electric current flowing into the electrolyte. The evidence outlined in this chapter for the photostimulation and charge injection properties suggests that porous plasmonic materials could have a central role in electrophysiology, biosensing and implantable device fields. These trials can be useful for further investigations of electron injection in liquids, leading to more efficient generation and exploitation of the use of plasmonic photostimulation *in vitro* and *in vivo* biological studies.

The measurements of photo-generated current of porous gold were found to be in agreement with the activity stimulation data collected from CMOS MEA presented in the previous section (*Optical stimulation of action potential in cardiomyocyte*). In fact, we demonstrate that we were able to elicit cardiomyocyte activity using the porous platinum of the CMOS electrodes. More in details, the laser excitation of the rough platinum can either stimulate a single secondary propagation pattern or promote temporarily the stimulated cell to the role of pacemaker, creating new artificial propagation pathways. Thus, this effect has a very promising potential because it offers the possibility to use the technique for cell stimulation at single cell level on high-density, high-resolution sensors, opening the way to new methodologies for studying cardiomyocytes and cardiac tissues.

In perspective, the current system has to be further investigated in order to develop a reliable and efficient protocol for optical cardiac stimulation[130]. The future developments, in combination with optical properties, sensing and recording techniques may lead to an integrated system for multimodal cells analysis and stimulation.

# Conclusions

In this work, I presented the Microfluidic Multielectrode array device able of simultaneous intra- and extra-cellular recording and selective intracellular delivery. By means of electroporation, the nanostructures are able to porate the cellular membrane allowing a way of access to the inner compartment of the cell. This method together with the flow-through design of the device, permitted to combine in the same platform optimal intracellular recording and highly localized delivery abilities.

The latter aspect was tested effectively injecting membrane-permeant and non-permeant molecules from the hollow nanochannels. After this, we successfully performed intracellular recording on human iPSC derived cardiomyocyte and HL-1 cells with the concurrent intracellular delivery on few selected cells. The platform proved efficient in acquiring signals from primary rat hippocampal and cortical neurons as well, discriminating the typical bursting activity of matured neuronal cultures. The recording was performed while stimulating molecules were injected into the culture from the hollow nanostructures, showing an effect on the firing activity only in neurons laying on the sites of delivery.

The platform, therefore, could be suitable for studying early stage neurologic pathologies by selectively induce the pathology only on few cells of a completely monitored network. Moreover, it could be used to evaluate the toxicity of pharmaceutical agents, drug-loaded nanocarrier or compounds with high spatial resolution. Therefore, this tool could allow to attain deeper knowledge in toxicological studies and therapeutics for neuro-related syndromes or cardiological conditions.

To further improve the impact of this platform, plasmonic metamaterials were explored as replacement of the 3D nanochannels in order to lower the cost of the above described platform and to lead to a multifunctional device for research use. As depicted in the schematic (fig. 22), the device design will be preserved, while the fabrication efficiency will be increased together with the number of electrodes for intra- and extra-cellular recording. The delivery capability of such a platform will be tested with different compound and molecules in order to reach an optimal reliability. The multi-purposes platform could be used for simultaneous non-destructive chemical analysis, such as in the case of SERS applications, for electrogenic cell activity recording, for cardiac cells eliciting and at the same time localized and intracellular drug delivery. Such a device will open the way for a more complete study of drug interactions and long term label free pathology development analysis and treatment enabling the possibility of observing and controlling cells interaction at bioelectrical and chemical level.

Moreover, the possibility of charge injection from plasmonic metamaterials was explored for plasmonic photostimulation of action potentials in *in vitro* cell cultures. The prospect to apply these materials for cell stimulation at single cell level on high-density and high-resolution commercial CMOS sensors was proven with the aim of studying cardiomyocytes and cardiac tissues. Further studies will have to be carried out in order to clearly define a protocol for cells stimulation with both platforms that are already on the market and still under development. Nevertheless, integration of cell stimulation technologies during screening assay would improve novel drugs development and pharmacological care enabling the evaluation of treatments effect on cells in different functional states [131]. The promising results achieved, in combination with optical properties, sensing and recording capabilities of plasmonic metamaterials, will lead to the development of combined systems for multimodal cells examination and stimulation.

# *Publications*

1. **G. Bruno**, N. Colistra, G. Melle, A. Cerea, F. De Angelis and M. Dipalo, *Microfluidic multielectrode arrays for spatially localized drug delivery and electrical recordings of primary neuronal cultures*, Front. Bioeng. Biotechnol.,(2020) https://doi.org/10.3389/fbioe.2020.00626

2. **G. Bruno**, Melle G., Barbaglia A., Iachetta G., Melikov R., Perrone M., Dipalo M., De Angelis F. *All-Optical and Label-Free Stimulation of Action Potentials in Neurons and Cardiomyocytes by Plasmonic Porous Metamaterials*, Advanced Science(2021),doi:10.1002/advs.202100627

3. A. Cerea, V. Caprettini, **G. Bruno**, L. Lovato, G. Melle, F. Tantussi, R. Capozza, F. Moia, M. Dipalo, F. De Angelis, *Selective intracellular delivery and intracellular recordings combined in MEA biosensors*, *Lab Chip* (2018), doi:10.1039/c8lc00435h.

4. Melle G., **Bruno G.**, Maccaferri N., Iachetta G., Colistra N., Barbaglia A., Dipalo M., De Angelis F.. *Intracellular Recording of Human Cardiac Action Potentials on Market-Available Multielectrode Array Platforms*, Frontiers in Bioengineering and Biotechnology(2020).

5. M. Dipalo, A. F. McGuire, H.-Y. Lou, V. Caprettini, G. Melle, **G. Bruno**, C. Lubrano, L. Matino, X. Li, F. De Angelis, B. Cui, F. Santoro, *Cells adhering to 3D vertical nanostructures: cell membrane reshaping without stable internalization*, *Nano Lett.* 18, acs.nanolett.8b03163 (2018).

6. M. Dipalo, G. Melle, L. Lovato, A. Jacassi, F. Santoro, V. Caprettini, A. Schirato, A. Alabastri, D. Garoli, **G. Bruno**, F. Tantussi, F. De Angelis, *Plasmonic meta-electrodes allow intracellular recordings at network level on high-density CMOS-multi-electrode arrays*, *Nat. Nanotechnol.* (2018), doi:10.1038/s41565-018-0222-z.

7. M. Dipalo, V. Caprettini, **G. Bruno**, F. Caliendo, L. D. Garma, G. Melle, M. Dukhinova, V. Siciliano, F. Santoro, F. De Angelis, *Membrane Poration Mechanisms at the Cell–Nanostructure Interface*, *Adv. Biosyst.* (2019), doi:10.1002/adbi.201900148.

8. A. Cerea, V. Caprettini, G. Melle, **G. Bruno**, M. Leoncini, A. Barbaglia, F. Santoro, M. Dipalo, *Coaxial-like three-dimensional nanoelectrodes for biological applications*, *Microelectron. Eng.* (2018), doi:10.1016/j.mee.2017.11.014.

9. M. Dipalo, A. F. McGuire, H. Y. Lou, V. Caprettini, G. Melle, **G. Bruno**, C. Lubrano, L. Matino, X. Li, F. De Angelis, B. Cui, F. Santoro, *Cells Adhering to 3D Vertical Nanostructures: Cell Membrane Reshaping without Stable Internalization*, *Nano Lett.* (2018), doi:10.1021/acs.nanolett.8b03163.

10. M. Ardini, J. A. Huang, C. S. Sánchez, M. Z. Mousavi, V. Caprettini, N. Maccaferri, G. Melle, **G. Bruno**, L. Pasquale, D. Garoli, F. De Angelis, *Live Intracellular Biorthogonal Imaging by Surface Enhanced Raman Spectroscopy using Alkyne-Silver Nanoparticles Clusters*, Sci. Rep. (2018), doi:10.1038/s41598-018-31165-3.